\documentclass[a4paper, 12pt, leqno]{scrartcl}

\let\originallhook\lhook
\usepackage{amsmath,amssymb,amsthm, bbding, cite, color, eurosym, extarrows, url,
bbm,stmaryrd,
mathrsfs, mathtools, mdframed, MnSymbol, 
yfonts,a4wide}
\usepackage[left=2cm,right=2cm,top=1.9cm,bottom=2.05cm]{geometry}
\let\lhook\originallhook
 
\DeclareMathOperator{\eq}{eq}

\DeclareMathOperator{\Loc}{\mathbf{Loc}}
\DeclareMathOperator{\Phys}{\mathbf{Phys}}

\DeclareMathOperator{\Sympl}{\mathbf{Sympl}}
\DeclareMathOperator{\pSympl}{\mathbf{pSympl}}

\DeclareMathOperator{\pSymplK}{\pSympl_\mathbb{K}}
\DeclareMathOperator{\mpSymplK}{\pSympl^m_\mathbb{K}}
\DeclareMathOperator{\SymplK}{\Sympl_\mathbb{K}}

\DeclareMathOperator{\pSymplC}{\pSympl_\mathbb{C}}
\DeclareMathOperator{\mpSymplC}{\pSympl^m_\mathbb{C}}

\DeclareMathOperator{\uAlg}{\mathbf{*\!Alg}_\mathbbm{1}}
\DeclareMathOperator{\umAlg}{\mathbf{*\!Alg}^m_\mathbbm{1}}

\DeclareMathOperator{\Aut}{Aut}

\DeclareMathOperator{\rce}{rce}
\DeclareMathOperator{\kin}{kin}
\DeclareMathOperator{\dyn}{dyn}
\DeclareMathOperator{\img}{img}
\DeclareMathOperator{\supp}{supp}
\DeclareMathOperator{\id}{id}

\DeclareMathOperator{\vol}{\,vol}
\DeclareMathOperator{\omegaa}{{\L\omega\R}}
\DeclareMathOperator{\etaa}{{\L\eta\R}}

\renewcommand{\L}{\left\llbracket}
\newcommand{\R}{\right\rrbracket}

\DeclareMathOperator{\iu}{i}

\DeclareMathOperator{\Euro}{\text{\euro}}

\newcommand{\Mb}{{\boldsymbol{M}}}
\newcommand{\Nb}{{\boldsymbol{N}}}
\DeclareMathOperator{\rad}{rad}
\newcommand{\ogth}{{\mathfrak o}}
\newcommand{\tgth}{{\mathfrak t}}
\newtheorem{thm}{Theorem}[section]
\newtheorem{prop}[thm]{Proposition}
\newtheorem{lem}[thm]{Lemma}
\newtheorem{cor}[thm]{Corollary}
\theoremstyle{definition}
\newtheorem{defn}[thm]{Definition}

\newcounter{mnotecount}[section]

\begin{document}

\title{Dynamical locality of the free \\ Maxwell field}
\author{Christopher J. Fewster\footnote{chris.fewster@york.ac.uk}\enspace and Benjamin Lang\footnote{bl620@york.ac.uk}\\
 Department of Mathematics, University of York,\\
 Heslington, York YO10 5DD, U.K.}
\date{\today}
\maketitle

\begin{abstract}
\noindent\textbf{Abstract:} We consider the non-interacting source-free Maxwell field, described both in terms of the vector potential and the field strength. Starting from the classical field theory on contractible globally hyperbolic spacetimes, we extend the classical field theory to general globally hyperbolic spacetimes in two ways to
obtain a `universal' theory and a `reduced' theory. The quantum field theory in terms of the unital $*$-algebra of the smeared quantum field is then obtained by an application of a suitable quantisation functor. 
We show that the universal theories fail local covariance and dynamical locality owing to the possibility of having non-trivial radicals in the classical and non-trivial centres in the quantum case. The reduced theories are both locally covariant and dynamically local. 
These models provide new examples relevant to the discussion of
how theories should be formulated so as to describe the 
same physics in all spacetimes.
\end{abstract}

\section{Introduction}
There have been a number of recent developments in the quantum field theory of electromagnetism, and other gauge field theories in the frameworks of algebraic quantum field theory \cite{HK64,LQP96} and locally covariant quantum field theory \cite{BFV03}. 
For example, the results for the initial value problem and the quantisation in \cite{Dim92} were generalised to differential $p$-form fields in \cite{Pf09}, Hadamard states were discussed in \cite{FePf03,DS13} and the Reeh-Schlieder property was analysed in \cite{D11}. However, these treatments have in common that they make some assumptions on the topology of the underlying spacetime. Approaches which do not make such assumptions are \cite{DL12}, which treats field strengths, \cite{DHS12}, which treats the vector potential, and \cite{FS13}, which discusses the Gupta-Bleuler formalism in curved spacetimes with the intention to couple the Dirac field with the electromagnetic field. A consideration of electromagnetism in the spirit of Yang-Mills gauge theories is given in the series of papers \cite{BDS12,BDS13,BDHS13}. The renormalisability of quantum Yang-Mills theories in curved spacetimes was established in \cite{Hol08} and a general setting for linear quantised gauge field theories is given in \cite{HS13}. One might also mention the progress made in linearised quantum gravity \cite{FH13}, which partly inspired some of the work just discussed. 

An interesting feature of the theories mentioned is that they do not conform to the
definition of locally covariant quantum field theories given in \cite{BFV03}. Among other things, this definition requires that, whenever a globally hyperbolic spacetime $\Mb$ is (suitably) embedded as a subspacetime of another such spacetime $\Nb$, the corresponding algebra of the quantum theory on $\Mb$ should be {\em injectively} mapped into the algebra of the theory on $\Nb$. However, 
the free electromagnetic field can support topological configurations labelled by de Rham cohomology 
classes, and as de Rham cohomology does not respect injectivity under spacetime embeddings, 
any theory that is sensitive to these topological configurations will fail to be locally covariant. 
As emphasised in \cite{DHS12}, it is the locality, rather than covariance, which is lost, as
the price for incorporating observables such as those related to Gauss' law. 

The aim of the present paper is to study the extent to which the quantised electromagnetic field has the property of {\em dynamical locality}. This notion was introduced in \cite{SPASs12} as an extra condition on locally covariant physical theories \cite{BFV03} and is closely related to the problem of formalising what it means for a theory to represent the same physics in all spacetimes (SPASs). It is of interest for other reasons too: for example,
as a key hypothesis for a no-go theorem concerning preferred states
in locally covariant quantum field theories~\cite[\S 6.3]{SPASs12}.
Dynamical locality has been established for a number
of theories, including the massive free scalar field \cite{FeV12}, the non-minimally coupled scalar field and the enlarged algebra of Wick polynomials \cite{Fer13}, the Dirac field \cite{Fer13b} and also the inhomogeneous Klein--Gordon theory \cite{FeSch14}; for a more detailed summary, see Section~\ref{sec conclusions}. While one may construct unphysical theories that fail dynamical locality~\cite{SPASs12}, 
the only `reasonable' theories known to do so (at least in four dimensions) are those containing the massless minimally coupled scalar field as a subtheory, for reasons that can be traced to a rigid gauge invariance. Electromagnetism, as a local gauge theory, evidently presents an interesting test case for dynamical locality. 

An immediate question is which of the various frameworks mentioned above should be used for this task. While the original formulation of 
dynamical locality concerns locally covariant theories obeying the timeslice axiom, one may modify the definition to apply to theories that do not respect injectivity, provided they nonetheless satisfy the timeslice axiom. 
We will study two models of electromagnetism, one of which respects
neither injectivity nor dynamical locality, while the other respects both. 
Each of these models can be formulated equivalently in terms of 
the field strength or vector potential, and both obey the timeslice axiom.  

At first sight, it may seem quixotic to formulate electromagnetism using
field strengths rather than vector potentials. However, with the exception
of \cite{FS13}, the literature on electromagnetism cited above focusses
entirely on gauge-invariant smearings of the vector potential, following the lead of \cite{Dim92}. In such models, the distinction between fields and potentials reduces to topological considerations; in fact
they all coincide if restricted to contractible spacetimes.  Our basic approach, following \cite{DL12}, is to take the theory on contractible
spacetimes and to ask how it may be extended to spacetimes with arbitrary topologies in a functorial way. This differs from other, more global, approaches like \cite{BDS12,BDS13,DHS12} insofar as we are led to our global theory (on non-contractible spacetimes) by local reasoning. Such an extension was already achieved in \cite{DL12} for the quantised free Maxwell field in terms of the field strength tensor using Fredenhagen's idea of the universal algebra
\cite{Fre90}. Theories obtained in this way will be called `universal' theories; as the field strength and vector potential formulations
of electromagnetism coincide in contractible spacetimes, their
corresponding universal theories are also equivalent: this is 
a generalisation of the ``natural algebraic relation'' described by \cite{Bong77} between the Borchers-Uhlmann algebras for the field strength description and the vector potential description of the quantum theory of the free Maxwell field in Minkowski space.

The universal theories do not obey local covariance: in spacetimes with non-trivial second de Rham cohomology, the corresponding pre-symplectic spaces (in the classical description) possess non-trivial radicals, while the corresponding $*$-algebras (in the quantised description) possess 
non-trivial centres. Such elements are lost under embeddings into
spacetimes with trivial cohomology. To remedy this, we also consider
`reduced' theories of electromagnetism which quotient out non-trivial radicals or centres -- similar ideas have been proposed in~\cite{DHS12,Benini2014}.  
 As we will show, the reduced theories are both
locally covariant (by design) and, which is not so obvious, dynamically local.

The paper is structured as follows. We begin with some preliminary work, collecting notions of dynamical locality in Section \ref{sec loc cov} and recalling some exterior calculus of differential forms in Section \ref{sec differential forms}. Next, we review the classical and the quantum field theory of the free Maxwell field in Section \ref{sec universal theory}. In doing so, we will also discuss the natural isomorphism between the field strength description and the vector potential description of the classical and the quantised theory of the free Maxwell field and also how 
electromagnetic duality is implemented in the theory.  In Section \ref{sec dynamical locality universal theory}, we will see that the classical and the quantised universal theory obtained in Section \ref{sec universal theory} fail local covariance and dynamical locality due to topological reasons 
already mentioned. This failure can be remedied, leading to a locally covariant and dynamically local {\em reduced theory} (classical and quantum) of the free Maxwell field, which will be the topic of Section \ref{sec dynamical locality reduced theory}. In Section~\ref{sec conclusions}, we 
discuss the status of dynamical locality, the 
categorical structure underlying some of our constructions, and the relation of our present
work to the discussions of SPASs in~\cite{SPASs12,FeV12}.


\section{Local covariance and dynamical locality}\label{sec loc cov}
We briefly review the framework of local covariance introduced in
\cite{BFV03}, in which physical theories are described as functors
between a category of spacetimes and a category of physical systems.
We also define the notion of dynamical locality~\cite{SPASs12}.
\subsection{Spacetimes and physical systems}
 
The category of spacetimes, $\Loc$, has as its objects all oriented globally hyperbolic spacetimes $\Mb=\left(M,g,\ogth,\tgth\right)$ of dimension $4$ and signature $\left(+,-,-,-\right)$, where $\ogth$ is the orientation and $\tgth$ is the time-orientation. A $\Loc$-morphism $\psi:\Mb\to\Nb$ is an isometric smooth embedding which preserves the orientation and the time-orientation and whose image $\psi\left(M\right)$ is causally convex\footnote{$\psi\left(M\right)$ is causally convex in $\Nb$ if and only if each causal smooth curve in $\Nb$ with endpoints in $\psi\left(M\right)$ is entirely contained in $\psi\left(M\right)$.} in $\Nb$. 

The physical systems under consideration shall form the objects of a category $\Phys$, so that a morphisms of $\Phys$ represents
an inclusion of one system as a subsystem of another. The category of  $\Phys$ is subjected to further conditions \cite[\S 3.1]{SPASs12}: to be specific it is required that all $\Phys$-morphisms are monic and that $\Phys$ has equalisers, intersections, unions\footnote{For the categorical notions of equalisers, which are also known as \emph{difference kernels}, intersections and unions see \cite{Par70} or \cite[Appx.B]{SPASs12}.} and an initial object, which represents the trivial physical theory.

We will consider just a few candidates for $\Phys$ in this paper:
\begin{enumerate}
\item[$\bullet$] $\umAlg$: $A\in\umAlg$ if and only if $A$ is a unital $*$-algebra over $\mathbb{C}$; for $A,B\in\umAlg$, $\varphi\in\umAlg\left(A,B\right)$ if and only if $\varphi:A\to B$ is a unital $*$-monomorphism.
\item[$\bullet$] $\mpSymplK$: Objects are (complexified if $\mathbb{K}=\mathbb{C}$) {\em pre-symplectic spaces}, $\left(V,\omega,C\right)$, where
 $V$ is a $\mathbb{K}$-vector space, $C$ a $C$-involution on $V$ (which is omitted or set to be the identity on $V$ if $\mathbb{K}=\mathbb{R}$)\footnote{A \emph{$C$-involution} on a complex vector space $V$ is a complex-conjugate linear map $C:V\longrightarrow V$ satisfying $C\circ C=\id_V$.} and $\omega$ a (possibly degenerate) skew-symmetric $\mathbb{K}$-bilinear form satisfying $\omega\circ\left(C\times C\right)=\overline{\phantom{n}}\circ\omega$; the morphisms
are symplectic $C$-monomorphisms, i.e., 
$f\in\mpSymplK\left(\left(V,\omega,C\right),\left(V',\omega',C'\right)\right)$ is
an injective $\mathbb{K}$-linear map $f:V\to V'$ such that $\omega'\circ\left(f\times f\right)=\omega$ and $f\circ C=C'\circ f$.
\end{enumerate}
We will also consider modifications of the categories mentioned so far as auxiliary structures.  $\Loc_\copyright$ is the full subcategory of $\Loc$ whose objects are contractible. $\uAlg$ is defined
in the same way as its subcategory $\umAlg$, but dropping the restriction of injectivity and allowing general unital $*$-homomorphisms. 
Similarly, $\pSymplK$ is defined in the same way as its subcategory $\mpSymplK$, dropping the restriction to injective morphisms.\footnote{Note that in \cite{FeV12}, $\pSymplK$ denotes the category we call $\mpSymplK$ here. As we will need to allow
for non-monic morphisms when considering the universal theory of the free Maxwell field, it is necessary to unambiguously indicate whether we only allow for monics or not.} Finally, $\SymplK$ is the full subcategory of $\mpSymplK$, where the (complexified if $\mathbb{K}=\mathbb{C}$) pre-symplectic form is now assumed to be weakly non-degenerate.      

\subsection{The relative Cauchy evolution}
We call a $\Loc$-morphism $\psi:\Mb\to\Nb$ \emph{Cauchy} whenever the image $\psi\left(M\right)$ contains a Cauchy surface for $\Nb$ (see \cite[Appx.A.1]{SPASs12} for some properties of Cauchy morphisms). A locally covariant theory (LCT) $\mathcal{A}:\Loc\to\Phys$ is said to obey the \emph{time-slice axiom} if and only if $\mathcal{A}\psi:\mathcal{A}\Mb\to\mathcal{A}\Nb$ is a $\Phys$-isomorphism whenever $\psi\in\Loc\left(\Mb,\Nb\right)$ is Cauchy.

For LCTs obeying the time-slice axiom, it is possible to define the relative Cauchy evolution \cite{BFV03}, which captures the dynamical reaction of the LCT to a local perturbation of the background metric; its
functional derivative with respect to the metric perturbation is 
closely related to the stress-energy tensor of the theory, see \cite{BFV03,SPASs12,FeV12}.

Let $\Mb=\left(M,g,\ogth,\tgth\right)\in\Loc$. A \emph{globally hyperbolic perturbation} $h$ of $\Mb$ is a compactly supported, symmetric and smooth tensor field such that the modification $\Mb\left[h\right]:=\left(M,g+h,\ogth,\tgth_{h}\right)$ becomes a $\Loc$-object, where $\tgth_{h}$ is the unique choice for a time-orientation on $\left(M,g+h\right)$ that coincides with $\tgth$ outside $\supp h$. We write $H\left(\Mb\right)$ for all globally hyperbolic perturbations of $\Mb$, while $H\left(\Mb;K\right)$ denotes the subset of all globally hyperbolic perturbations whose support is contained in a subset $K\subseteq M$. For each $h\in H\left(\Mb\right)$, we define open sets $M^\pm\left[h\right]:=M\setminus J_\Mb^{\mp}\left(\supp h\right)$, which will become $\Loc$-objects in their own right if endowed with the structures induced by $\Mb$ or $\Mb\left[h\right]$\footnote{It does not matter whether we use $\Mb$ or $\Mb\left[h\right]$ since $M^\pm\left[h\right]\cap\supp h=\emptyset$.} by \cite[Lem.3.2(a)]{SPASs12}. We denote these $\Loc$-objects by $\Mb^\pm\left[h\right]=\Mb|_{M^\pm\left[h\right]}=(M^{\pm}\left[h\right],g|_{M^\pm\left[h\right]},\ogth|_{M^\pm\left[h\right]},\tgth|_{M^\pm\left[h\right]})$. By \cite[Lem3.2(b)]{SPASs12}, the inclusion maps
\begin{align*}
\iota_{M^{\pm}\left[h\right]M}:M^{\pm}\left[h\right]\longrightarrow M&&\text{and}&&\iota_{M^{\pm}\left[h\right]M\left[h\right]}:M^{\pm}\left[h\right]\longrightarrow M\left[h\right]
\end{align*}
become Cauchy morphisms, which we will denote by 
\begin{align*}
\imath^\pm_\Mb\left[h\right]:\Mb^\pm\left[h\right]\longrightarrow\Mb&&\text{and}&&\jmath^\pm_\Mb\left[h\right]:\Mb^\pm\left[h\right]\longrightarrow\Mb\left[h\right].
\end{align*}
Now, given a LCT $\mathcal{A}:\Loc\to\Phys$ which obeys the time-slice axiom, the relative Cauchy evolution for $\mathcal{A}$ induced by $h\in H\left(\Mb\right)$ is the $\Phys$-automorphism $\mathcal{A}\Mb\to\mathcal{A}\Mb$ defined by
\begin{align}\label{relative Cauchy evolution}
\rce^\mathcal{A}_\Mb\left[h\right]:=\mathcal{A}\left(\imath^-_\Mb\left[h\right]\right)\circ\left(\mathcal{A}\left(\jmath^-_\Mb\left[h\right]\right)\right)^{-1}\circ\mathcal{A}\left(\jmath^+_\Mb\left[h\right]\right) \circ\left(\mathcal{A}\left(\imath^+_\Mb\left[h\right]\right)\right)^{-1}.
\end{align}

\subsection{The dynamical net and dynamical locality}\label{subsec dynamical net}
For $\Mb\in\Loc$, let $\mathscr{O}\left(\Mb\right)$ denote the set of all open globally hyperbolic subsets of $\Mb$. If $\mathcal{A}:\Loc\to\Phys$ is a LCT, then the \emph{kinematic net} of $\mathcal{A}$ for $\Mb$ is defined by the rule $O\mapsto\left(\mathcal{A}\iota_\mathcal{O}:\Mb|_O\to\mathcal{A}\Mb\right)$, as $O$ ranges over the nonempty elements of $\mathscr{O}\left(\Mb\right)$, where $\iota_O:O\to M$ denotes the inclusion map. In this context, $\mathcal{A}\Mb|_O$ is also denoted by $\mathcal{A}^{\kin}\left(\Mb;O\right)$ and $\mathcal{A}\iota_O$ by $\alpha^{\kin}_{\Mb;O}$.
 
The definition of the {\em dynamical net} for a LCT $\mathcal{A}:\Loc\to\Phys$ obeying the time-slice axiom consists of three steps~\cite{SPASs12}: first, take $K$ compact in $\Mb\in\Loc$ and consider all elements of $\mathcal{A}\Mb$ insensitive to a globally hyperbolic perturbation $h\in H\left(\Mb;K^\perp\right)$, where $K^\perp:=M\setminus J_\Mb(K)$. In categorical terms, we are looking at the equaliser
\begin{align*}
\eq\left(\rce^\mathcal{A}_\Mb\left[h\right],\id_{\mathcal{A}\Mb}\right):E\left(\rce^\mathcal{A}_\Mb\left[h\right],\id_{\mathcal{A}\Mb}\right)\longrightarrow\mathcal{A}\Mb.
\end{align*}
In $\Phys=\mpSymplK$ or $\Phys=\umAlg$ this equaliser is just the inclusion of the (complexified if $\mathbb{K}=\mathbb{C}$) pre-symplectic subspace or unital $*$-subalgebra with the underlying set $\left\{a\in\mathcal{A}\Mb\mid\rce^\mathcal{A}_\Mb\left[h\right] a=a\right\}$ into $\mathcal{A}\Mb$. In the second step, 
we isolate those elements of $\mathcal{A}\Mb$ that are insensitive to all globally hyperbolic perturbations $h\in H\left(\Mb;K^\perp\right)$
supported in the region that is causally inaccessible to $K$. This can be used to define what it means for an observable to be localised in $K$. In categorical terms, we form the intersection
\begin{align*}
\bigwedge_{h\in H\left(\Mb;K^\perp\right)}\eq\left(\rce^\mathcal{A}_\Mb\left[h\right],\id_{\mathcal{A}\Mb}\right):\bigwedge_{h\in H\left(\Mb;K^\perp\right)}E\left(\rce^\mathcal{A}_\Mb\left[h\right],\id_{\mathcal{A}\Mb}\right)\longrightarrow\mathcal{A}\Mb.
\end{align*}
We will write
\begin{equation*}
\alpha^\bullet_{\Mb;K} \!:=\!\bigwedge_{h\in H\left(\Mb;K^\perp\right)}\eq\left(\rce^\mathcal{A}_\Mb\left[h\right],\id_{\mathcal{A}\Mb}\right), \qquad\text{and}\qquad \mathcal{A}^{\bullet}\left(\Mb;K\right):=\bigwedge_{h\in H\left(\Mb;K^\perp\right)}E\left(\rce^\mathcal{A}_\Mb\left[h\right],\id_{\mathcal{A}\Mb}\right)
\end{equation*}
for convenience. For $\Phys=\mpSymplK$ or $\Phys=\umAlg$,
we may identify $\mathcal{A}^{\bullet}\left(\Mb;K\right)= \left\{a\in\mathcal{A}\Mb\mid\rce^\mathcal{A}_\Mb\left[h\right] a=a \enspace\forall h\in H\left(\Mb;K^\perp\right)\right\}$, and 
$\alpha^\bullet_{\Mb;K}$ as the inclusion of $\mathcal{A}^{\bullet}\left(\Mb;K\right)$ as a  (complexified if $\mathbb{K}=\mathbb{C}$)  pre-symplectic subspace or unital $*$-subalgebra of $\mathcal{A}\Mb$. 
Thirdly and finally, we consider for $O\in\mathscr{O}\left(\Mb\right)$ the union
\begin{align*}
\bigvee_{K\in\mathcal{K}\left(\Mb;O\right)}\alpha^\bullet_{\Mb;K}:\bigvee_{K\in\mathcal{K}\left(\Mb;O\right)}\mathcal{A}^\bullet\left(\Mb;K\right)\longrightarrow\mathcal{A}\Mb,
\end{align*}
where $\mathscr{K}\left(\Mb;O\right)$ is the set of all compact subsets of $O\in\mathscr{O}\left(\Mb\right)$ which have a \emph{multi-diamond} open neighbourhood whose \emph{base} is contained in $O$. That is, each $K\in\mathscr{K}\left(\Mb;O\right)$ has an open neighbourhood which is the union of finitely many, causally disjoint and open sets of the form $D_\Mb\left(B\right)$, where $D_\Mb$ denotes the Cauchy development in $\Mb$ and $B\subseteq O$ is a Cauchy ball; that is, 
$B$ is an open set of a smooth spacelike Cauchy surface $\Sigma$ for $\Mb$ diffeomorphic to an open ball of $\mathbb{R}^3$ under a smooth chart for $\Sigma$. For this choice of $\mathscr{K}\left(\Mb;O\right)$, we refer the reader to \cite[\S 5]{SPASs12}.
Note that every point $x\in\Mb$ is contained in a Cauchy ball: let $\Sigma_x$
be any smooth spacelike Cauchy surface for $\Mb$ containing $x$, choose any 
smooth chart $\varphi:U\to W\subseteq\mathbb{R}^3$ for $\Sigma_x$ with $x\in U$ and $\varepsilon>0$ such that the $\varepsilon$-ball around $\varphi\left(x\right)$ is contained in $W$, and then take $B_x:=\varphi^{-1}\left(B_\delta(\varphi\left(x\right))\right)$ with $\delta<\varepsilon$.
  For the sake of convenience, we set $\alpha^{\dyn}_{\Mb;O}:=\bigvee_{K\in\mathcal{K}\left(\Mb;O\right)}\alpha^\bullet_{\Mb;K}$ and $\mathcal{A}^{\dyn}_{\Mb;O}:=\bigvee_{K\in\mathcal{K}\left(\Mb;O\right)}\mathcal{A}^\bullet_{\Mb;K}$ for all $O\in\mathscr{O}\left(\Mb\right)$. For $\Phys=\mpSymplK$ or $\Phys=\umAlg$, $\alpha^{\dyn}_{\Mb;O}$ is the inclusion of the (complexified if $\mathbb{K}=\mathbb{C}$) pre-symplectic subspace or unital *-subalgebra $\mathcal{A}^{\dyn}_{\Mb;O}$ generated by $\bigcup_{K\in\mathscr{K}\left(\Mb;O\right)}\mathcal{A}^{\bullet}\left(\Mb;K\right)$. The rule $\mathscr{O}\left(\Mb\right)\ni O\longmapsto\alpha^{\dyn}_{\Mb;O}$ is called the \emph{dynamical net} of $\mathcal{A}$ for $\Mb$ and a LCT $\mathcal{A}:\Loc\to\Phys$ obeying the time-slice axiom is called \emph{dynamically local} if and only if the kinematic and the dynamical net are equivalent in the sense of subobjects (\!\!\cite[\S 1.6]{Par70}, \cite[Appx.B]{FeV12}).

\section{Some preliminaries on differential forms}\label{sec differential forms}
Differential forms allow for an elegant geometrical description of electromagnetism, that extends to curved spacetimes and allows for a relatively easy quantisation. 
For $\Mb\in\Loc$, we denote the $\mathcal{C}^\infty\left(M,\mathbb{K}\right)$-module of all $\mathbb{K}$-valued differential $p$-forms ($p\geq0$) by $\Omega^p\left(M;\mathbb{K}\right)$. Adding the subscript `0', i.e. writing $\Omega^p_0\left(M;\mathbb{K}\right)$, will denote the $\mathcal{C}^\infty\left(M,\mathbb{K}\right)$-module of all $\mathbb{K}$-valued differential $p$-forms of compact support. By convention, $\Omega^{-1}_{\left(0\right)}\left(M;\mathbb{K}\right)$ is
the trivial $\mathbb{K}$-vector space. 

Several operators on differential forms will be of importance to us. First, the \emph{exterior derivative}\footnote{The subscript `$(0)$' indicates that the map is well-defined for both with and without the subscript.} $d_\Mb:\Omega^p_{\left(0\right)}\left(M;\mathbb{K}\right)\to\Omega^{p+1}_{\left(0\right)}\left(M;\mathbb{K}\right)$ is given, in abstract index notation, by
\begin{flalign*}
&&\left(d_\Mb\omega\right)_{a_1\dots a_{p+1}}=\sum_{i=1}^{p+1}\left(-1\right)^{i+1}\nabla\!_{a_i}\,\omega_{a_1\dots a_{i-1}a_{i+1}\dots a_{p+1}},&&\makebox[0pt][r]{$\omega\in\Omega^{p}\left(M;\mathbb{K}\right)$,}
\end{flalign*}  
where $\nabla$ denotes the Levi-Civita connection on $\Mb$; by convention $d_\Mb:\Omega^{-1}_{\left(0\right)}\left(M;\mathbb{K}\right)\to \Omega^{0}_{\left(0\right)}\left(M;\mathbb{K}\right)$ is the zero map. The $\mathbb{K}$-vector space of all (compactly supported) $\mathbb{K}$-valued differential $p$-forms $\omega\in\Omega_{\left(0\right)}^p\left(M;\mathbb{K}\right)$ which are \emph{closed}, that is, $d_\Mb\omega=0$, is denoted by $\Omega^p_{(0),d}\left(M;\mathbb{K}\right)$. We say that $\omega\in\Omega^p_d\left(M;\mathbb{K}\right)$ is \emph{exact} if and only if there is $\theta\in\Omega^{p-1}\left(M;\mathbb{K}\right)$ such that $\omega=d_\Mb\theta$. For $p\geq 0$, the \emph{(compactly supported) de Rham cohomology groups} $H^p_{dR,\left(c\right)}\left(M;\mathbb{K}\right):=\Omega^p_{\left(0\right),d}\left(M;\mathbb{K}\right)/d_\Mb\Omega^{p-1}_{\left(0\right)}\left(M;\mathbb{K}\right)$ indicate to what extent the closed differential forms of a smooth manifold fail to be exact and are deeply connected to the topology of the manifold via singular homology. By Poincar\'{e} duality \cite[\S V.4]{Greub1}, we have $H^p_{dR}\left(M;\mathbb{K}\right)\cong\left(H^{4-p}_{dR,c}\left(M;\mathbb{K}\right)\right)^*$, where `$*$' denotes the dual. 

Next, the \emph{Hodge-$*$-operator} $*_\Mb:\Omega^p_{\left(0\right)}\left(M;\mathbb{K}\right)\to\Omega^{4-p}_{\left(0\right)}\left(M;\mathbb{K}\right)$ is the $\mathcal{C}^\infty\left(M,\mathbb{K}\right)$-module isomorphism defined by 
\begin{flalign*}
&&\omega\wedge*_\Mb\eta=\frac{1}{p!}\omega_{a_1\dots a_p}\eta^{a_1\dots a_p}\vol_\Mb,&&\makebox[0pt][r]{$\omega,\eta\in\Omega^p\left(M;\mathbb{K}\right)$,}
\end{flalign*} 
with inverse  $*_\Mb^{-1}=(-1)^{p(4-p)+1}*_\Mb$. The Hodge-$*$
provides a weakly non-degenerate $\mathbb{K}$-bilinear pairing $\int_M\left(\cdot\right)\wedge*_\Mb\left(\cdot\right)$ of  
$\Omega^p\left(M;\mathbb{K}\right)$ and $\Omega^p_0\left(M;\mathbb{K}\right)$.

Using the exterior derivative and the Hodge-$*$, we construct the \emph{exterior co\-derivative} $\delta_\Mb:=(-1)^p*_\Mb^{-1}d_\Mb*_\Mb:\Omega^p_{\left(0\right)}\left(M;\mathbb{K}\right)\to\Omega^{p-1}_{\left(0\right)}\left(M;\mathbb{K}\right)$, which is formally adjoint to $d_\Mb$ in the sense that
\begin{align*}
\int_M\omega\wedge*_\Mb\delta_\Mb\eta=\int_Md_\Mb\omega\wedge*_\Mb\eta
\end{align*}
whenever $\omega\in\Omega^p\left(M;\mathbb{K}\right)$ and $\eta\in\Omega^{p+1}\left(M;\mathbb{K}\right)$ such that $\supp\omega\cap\supp\eta$ is compact.  In abstract index notation
\begin{flalign*}
&&\left(\delta_\Mb\omega\right)_{a_1\dots a_{p-1}}=-\nabla\!_{a_0}\omega^{a_0}_{\phantom{a_0}a_1\dots a_{p-1}},&&\makebox[0pt][r]{$\omega\in\Omega^{p}\left(M;\mathbb{K}\right)$}.
\end{flalign*}
$\Omega^p_{\left(0\right),\delta}\left(M;\mathbb{K}\right)$ will denote the $\mathbb{K}$-vector space of all (compactly supported) $\mathbb{K}$-valued differential $p$-forms $\omega\in\Omega^p\left(M;\mathbb{K}\right)$ which are \emph{coclosed}, that is $\delta_\Mb\omega=0$. $\omega\in\Omega^p_\delta\left(M;\mathbb{K}\right)$ is called \emph{coexact} if and only if there is $\eta\in\Omega^{p+1}\left(M;\mathbb{K}\right)$ with $\omega=\delta_\Mb\eta$. Closed and coclosed as well as exact and coexact differential forms are related to each other by the Hodge-$*$-operator. 

The \emph{d'Alembertian} or \emph{wave operator} $\Box_\Mb:\Omega^p_{\left(0\right)}\left(M;\mathbb{K}\right)\to\Omega^p_{\left(0\right)}\left(M;\mathbb{K}\right)$
is defined by $\Box_\Mb:=-\delta_\Mb d_\Mb-d_\Mb\delta_\Mb$. In abstract index notation we have
\begin{flalign*}
&&\left(\Box_\Mb\omega\right)_{a_1\dots a_p}=g^{ab}\nabla\!_a\nabla\!_b\,\omega_{a_1\dots a_p}+\sum_{i=1}^p(-1)^pg^{ab}\,[\nabla\!_a,\nabla\!_{a_i}]\,\omega_{ba_1\dots a_{i-1}a_{i+1}\dots a_p},&&\omega\in\Omega^p\left(M;\mathbb{K}\right),
\end{flalign*}
which establishes that $\Box_\Mb$ is a normally hyperbolic differential operator of metric type (see \cite[\S 1.5]{BGP07} for a definition but note that \cite{BGP07} employ the $\left(-,+,+,+\right)$-metric signature). Hence, \cite{BGP07} shows that $\Box_\Mb$ has a well-posed Cauchy problem and that there are unique retarded and advanced Green's operators $G^\text{ret/adv}_\Mb$ such that
$\supp G_\Mb^\text{ret/adv}\omega\subseteq J^{+/-}_\Mb\left(\supp\omega\right)$ 
(usage of `advanced' and `retarded' is reversed in \cite{BGP07}). 
We will make extensive use of the difference  $G_\Mb:=G_\Mb^\text{ret}-G_\Mb^\text{adv}$.\footnote{Note that the references \cite{FePf03,FeV12} focus on the advanced minus retarded Green's operator.}  We collect some useful properties: 
\begin{lem} \label{lem:propsofG} The following hold for any $p\ge 0$:
(a) The identities
$G_\Mb d_\Mb\omega=d_\Mb G_\Mb\omega$ and $G_\Mb\delta_\Mb\omega=\delta_\Mb G_\Mb\omega$ hold for all $\omega\in\Omega^p_0\left(M;\mathbb{K}\right)$. 
(b) The kernel of $\Box_\Mb$ on $\Omega^p_0\left(M;\mathbb{K}\right)$ is trivial, while the
range of $G_\Mb$ on $\Omega^p_0\left(M;\mathbb{K}\right)$ coincides with the space of $\eta\in\Omega^p\left(M;\mathbb{K}\right)$ such that
$\Box_\Mb\eta=0$ and so that $\eta$ has spacelike compact support
(which is equivalent to having compact support on  
Cauchy surfaces~\cite{San13}). The kernel of $G_\Mb$ on $\Omega^p_0\left(M;\mathbb{K}\right)$ is given by $\Box_\Mb\Omega^p_0\left(M;\mathbb{K}\right)$.
(c) The identity $G_\Mb d_\Mb\delta_\Mb\omega = -
G_\Mb\delta_\Mb d_\Mb\omega$ holds for $\omega\in\Omega^p_0\left(M;\mathbb{K}\right)$.
(d) The kernels of $d_\Mb\Box_\Mb$ and $\delta_\Mb\Box_\Mb$ on $\Omega^p_0\left(M;\mathbb{K}\right)$ are $\Omega^p_{0,d}\left(M;\mathbb{K}\right)$ and $\Omega^p_{0,\delta}\left(M;\mathbb{K}\right)$ respectively. (e)  
The kernels of $d_\Mb G_\Mb \delta_\Mb$ and $\delta_\Mb G_\Mb d_\Mb$ on $\Omega^p_0\left(M;\mathbb{K}\right)$
are both equal to $\Omega^p_{0,d}\left(M;\mathbb{K}\right)\oplus\Omega^p_{0,\delta}\left(M;\mathbb{K}\right)$. 
\end{lem}
\noindent\textbf{\textit{Proof:}}  (a) is proved, e.g.,  in  \cite[Prop.2.1]{Pf09}; (b) is standard for normally hyperbolic operators, e.g., 
\cite[Thm.~3.4.7]{BGP07}; (c) is a special case of (b) using the definition of $\Box_\Mb$. For (d), we observe
that $d_\Mb\Box_\Mb\alpha=0$ for $\alpha\in\Omega^p_0\left(M;\mathbb{K}\right)$ implies $\Box_\Mb d_\Mb\alpha=0$ and hence that $d_\Mb\alpha=0$ by (b); conversely, it is clear that $\alpha\in\Omega^p_{0,d}\left(M;\mathbb{K}\right)$ implies  $d_\Mb\Box_\Mb\alpha=0$. Similarly, $\delta_\Mb\Box_\Mb\alpha=0$ if and only if $\delta_\Mb\alpha=0$.
Finally, if $d_\Mb G_\Mb \delta_\Mb\omega=0$ for $\omega\in\Omega^p_0\left(M;\mathbb{K}\right)$ then we also have $G_\Mb d_\Mb \delta_\Mb\omega=0$ and hence $d_\Mb \delta_\Mb\omega=\Box_\Mb\alpha$ for some $\alpha\in\Omega^p_0\left(M;\mathbb{K}\right)$ by (b); as it is clear that $d_\Mb\Box_\Mb\alpha=0$, 
(d) gives $\alpha\in\Omega^p_{0,d}\left(M;\mathbb{K}\right)$. By (c), we also have $G_\Mb \delta_\Mb d_\Mb\omega=0$  and by similar arguments, $\delta_\Mb d_\Mb\omega=\Box_\Mb\beta$ for $\beta\in\Omega^p_{0,\delta}\left(M;\mathbb{K}\right)$. We deduce that $\Box_\Mb(\omega+\alpha+\beta)=0$
and hence $\omega\in \Omega^p_{0,d}\left(M;\mathbb{K}\right)+\Omega^p_{0,\delta}\left(M;\mathbb{K}\right)$. This is actually a direct sum, because any  $\omega\in \Omega^p_{0,d}\left(M;\mathbb{K}\right)\cap \Omega^p_{0,\delta}\left(M;\mathbb{K}\right)$ obeys $\Box_\Mb\omega=0$, so the
intersection is trivial. The reverse inclusion is easily shown using (c).
 \hfill\SquareCastShadowTopRight


\section{Classical and quantum Maxwell theories}\label{sec universal theory}

\subsection{The initial value problem}
For $\Mb\in\Loc$, the free Maxwell equations for the electromagnetic field strength tensor $F\in\Omega^2\left(M;\mathbb{K}\right)$ are
\begin{align}\label{eq. free Maxwell F}
&&d_\Mb F=0&&\text{and}&&\delta_\Mb F=0.&&
\end{align}
Given the electric field $E\in\Omega^1_{0,\delta}\left(\Sigma;\mathbb{K}\right)$ and the magnetic field $B\in\Omega^2_{0,d}\left(\Sigma;\mathbb{K}\right)$ on a smooth spacelike Cauchy surface $\Sigma$ for $\Mb$ with inclusion map $\iota_\Sigma:\Sigma\to M$, we can formulate the well-posed initial value problem \cite[Prop.2.1]{DL12}:
\begin{align}\label{initial value problem F}
d_\Mb F=0,&&\delta_\Mb F=0,&&-\iota_\Sigma^*F=B&&\text{and}&&*_\Sigma\,\iota_\Sigma^**^{-1}_\Mb F=E.
\end{align}
Following \cite{Bong77}, we will generally call this the {\em F-description} of the free Maxwell field. 

As is well-known, on any $\Mb\in\Loc_\copyright$, every solution of (\ref{eq. free Maxwell F}) can be expressed in terms of a vector potential as $F=d_\Mb A$ (i.e., $F_{ab}=\nabla\!_a A_b-\nabla\!_b A_a$) because $H^2_{dR}\left(M;\mathbb{K}\right)=0$, whereupon the free Maxwell equations (\ref{eq. free Maxwell F}) can be re-expressed as the single equation $\delta_\Mb d_\Mb A=0$ 
for the electromagnetic vector potential $A\in\Omega^1\left(M;\mathbb{K}\right)$. Owing to gauge freedom, however, 
the initial value problem
\begin{align*}
\delta_\Mb d_\Mb A=0,&&-\iota_\Sigma^*A=\mathcal{A}&&\text{and}&&*_\Sigma\,\iota_\Sigma^**^{-1}_\Mb d_\Mb A=E,
\end{align*}
where $\Sigma$, $\iota_\Sigma$ and $E$ as above and $\mathcal{A}\in\Omega^1_0\left(\Sigma;\mathbb{K}\right)$ is the magnetic vector potential, i.e. $d_\Sigma\mathcal{A}=B$, is not well-posed. Instead, a well-posed initial value problem is obtained by passing to suitable equivalence classes of initial data and solutions \cite{Dim92,Pf09,DHS12}. We will generally refer to the description in terms of the vector potential as the {\em A-description} of the free Maxwell field.

\subsection{A classical phase space for contractible spacetimes}\label{subsec contractible classical phase space}
We continue to assume that $\Mb\in\Loc_\copyright$. In the F- and the A-description, there are three descriptions of the classical field theory in terms of a (possibly complexified) symplectic space: the phase space of the Cauchy data, the phase space of the solutions and the phase space of the test forms (cf. \cite[\S 3]{Dim92} for the case of the electromagnetic vector potential). However, these three choices are symplectomorphic and hence equivalent. We will find it convenient to work with the phase space of test forms, which we now discuss briefly.

As shown in the proof of \cite[Prop.2.1]{DL12}, any solution of (\ref{initial value problem F}) with compact support on Cauchy surfaces is also a solution for the initial value problem of the wave equation $\Box_\Mb F=0$ with compactly supported Cauchy data, and can be written as \cite[Prop.2.2]{DL12}: 
\begin{flalign*}
&&F=G_\Mb\left(d_\Mb\theta+\delta_\Mb\eta\right),&&\theta\in\Omega^1_{0,\delta}\left(M;\mathbb{K}\right),\,\eta\in\Omega^3_{0,d}\left(M;\mathbb{K}\right).
\end{flalign*}
This general form may be simplified as $\Mb$ is contractible (so $H^1_{dR}\left(M;\mathbb{K}\right)$ is trivial), and hence $\Omega^1_{0,\delta}\left(M;\mathbb{K}\right)=\delta_\Mb\Omega^2_0\left(M;\mathbb{K}\right)$ and $\Omega^3_{0,d}\left(M;\mathbb{K}\right)=d_\Mb\Omega^2_0\left(M;\mathbb{K}\right)$.  Making use of Lem.~\ref{lem:propsofG}, we see that any solution of (\ref{initial value problem F}) with compact support on Cauchy surfaces can be written 
\begin{flalign*}
&&F=d_\Mb G_\Mb\delta_\Mb\omega,&&\makebox[0pt][r]{$\omega\in\Omega^2_0\left(M;\mathbb{K}\right)$.}
\end{flalign*}
By Lem.~\ref{lem:propsofG}(e), $\omega,\eta\in\Omega^2_0(M;\mathbb{K})$ give rise to the same solution if and only if they differ by an element of $\Omega^2_{0,d}\left(M;\mathbb{K}\right)\oplus \Omega^2_{0,\delta}\left(M;\mathbb{K}\right)$. As $\Mb$ is contractible, we have
$\Omega^2_{0,d}\left(M;\mathbb{K}\right)=d_\Mb\Omega^1_0\left(M;\mathbb{K}\right)$ and  $\Omega^2_{0,\delta}\left(M;\mathbb{K}\right)=\delta_\Mb\Omega^3_0\left(M;\mathbb{K}\right)$, so the space of solutions may be described
as a (complexified if $\mathbb{K}=\mathbb{C}$) symplectic space  $\mathcal{F}\Mb:=\left(\left[\Omega^2_0\left(M;\mathbb{K}\right)\right],\textswab{w}_\Mb,\overline{\phantom{n}}\right)$,\footnote{The use of the same symbol $\mathcal{F}$ [and later $\mathcal{A}$] for both $\mathbb{K}=\mathbb{R}$ and $\mathbb{K}=\mathbb{C}$, should not give rise to any confusion.} where
\begin{align}\label{F-classic}
\begin{aligned}
\left[\Omega^2_0\left(M;\mathbb{K}\right)\right]:=\Omega^2_0\left(M;\mathbb{K}\right)\big/\left(d_\Mb\Omega^1_0\left(M;\mathbb{K}\right)\oplus\delta_\Mb\Omega^3_0\left(M;\mathbb{K}\right)\right),&&\\
\textswab{w}_\Mb\left(\left[\omega\right],\left[\eta\right]\right):=-\int_MG_{\Mb}\delta_\Mb\omega\wedge*_\Mb\delta_\Mb\eta,\quad \overline{\left[\omega\right]}:=\left[\overline{\omega}\right],&&
\left[\omega\right],\left[\eta\right]\in\left[\Omega^2_0\left(M;\mathbb{K}\right)\right],
\end{aligned}
\end{align}
(the complex conjugation is to be omitted if $\mathbb{K}=\mathbb{R}$).
For future reference, we observe that $\left[\Box_\Mb\Omega^2_0\left(M;\mathbb{K}\right)\right]=\left\{\left[0\right]\right\}$.
The fact that $\textswab{w}_\Mb$ is a well-defined and non-degenerate follows immediately from the following
result. 
\begin{lem}\label{lem ur sym} Let $\Mb\in\Loc$ (contractibility is not assumed). Then
\begin{align*}
\left(\omega ,\eta\right) &\mapsto -\int_MG_{\Mb}\delta_\Mb\omega\wedge*_\Mb\delta_\Mb\eta, 
\end{align*}
is a skew-symmetric, $\mathbb{K}$-bilinear form on $\Omega^2_0\left(M;\mathbb{K}\right)$, 
with radical  $\Omega^2_{0,d}\left(M;\mathbb{K}\right)\oplus \Omega^2_{0,\delta}\left(M;\mathbb{K}\right)$.
\end{lem}
\noindent\textbf{\textit{Proof:}} Bilinearity is obvious and skew-symmetry follows from general
properties of $G_\Mb$. Fixing $\omega \in\Omega^2_0\left(M;\mathbb{K}\right)$ and noting that
\begin{flalign}\label{eq:Gints}
&&\int_M G_\Mb\delta_\Mb\omega\wedge*_\Mb \delta_\Mb\eta &= \int_M 
d_\Mb G_\Mb\delta_\Mb\omega\wedge*_\Mb\eta&&\forall\eta\in\Omega^2_0\left(M;\mathbb{K}\right),
\end{flalign}
the non-degeneracy of the pairing $\int_M\left(\cdot\right)\wedge*_\Mb\left(\cdot\right):\Omega^2\left(M;\mathbb{K}\right)\times\Omega_0^2\left(M;\mathbb{K}\right)\to\mathbb{K}$ implies that the
left-hand side of \eqref{eq:Gints} vanishes for 
all $\eta\in\Omega^2_0\left(M;\mathbb{K}\right)$ if and only if $d_\Mb G_\Mb\delta_\Mb\omega=0$
and hence $\omega\in \Omega^2_{0,d}\left(M;\mathbb{K}\right)\oplus \Omega^2_{0,\delta}\left(M;\mathbb{K}\right)$ by Lem.~\ref{lem:propsofG}(e). 
\hfill\SquareCastShadowTopRight\par\bigskip

In the A-description, the classical field theory can be described by the (complexified if $\mathbb{K}=\mathbb{C}$) symplectic space $\mathcal{A}\Mb=\left(\left[\delta_\Mb\Omega^2_0\left(M;\mathbb{K}\right)\right], \textswab{v}_\Mb, \overline{\phantom{n}}\right)$, where 
(omitting the complex conjugation if $\mathbb{K}=\mathbb{R}$)
\begin{align}\label{A-classic}
\begin{aligned}
\left[\delta_\Mb\Omega^2_0\left(M;\mathbb{K}\right)\right]:=\delta_\Mb\Omega^2_0\left(M;\mathbb{K}\right)\big/\delta_\Mb d_\Mb\Omega^1_0\left(M;\mathbb{K}\right),&&\\
\textswab{v}_\Mb\left(\left[\theta\right],\left[\phi\right]\right):=-\int_MG_{\Mb}\theta\wedge*_\Mb\phi,\qquad
\overline{\left[\theta\right]}:=\left[\overline{\theta}\right],
&&\quad\left[\theta\right],\left[\phi\right]\in\left[\delta_\Mb\Omega^2_0\left(M;\mathbb{K}\right)\right],
\end{aligned}
\end{align} 
see \cite{Dim92,Pf09,D11,DS13}. Note, the first two references assume that $\Mb$ has compact Cauchy surfaces. This assumption is not necessary here (though we have contractibility at present). Also, recall the identity $\delta_\Mb\Omega^2_0\left(M;\mathbb{K}\right)=\Omega^1_{0,\delta}\left(M;\mathbb{K}\right)$ due to the assumption $\Mb\in\Loc_\copyright$.

Using the pushforward of compactly supported $\mathbb{K}$-valued differential forms, we obtain $\SymplK$-morphisms $\mathcal{F}\psi:\mathcal{F}\Mb\to\mathcal{F}\Nb$ and $\mathcal{A}\psi:\mathcal{A}\Mb\to\mathcal{A}\Nb$ by the definitions $\mathcal{F}\psi\left[\omega\right]:=\left[\psi_*\omega\right]$, $\omega\in\Omega^2_0\left(M;\mathbb{K}\right)$, and $\mathcal{A}\psi\left[\theta\right]:=\left[\psi_*\theta\right]$, $\theta\in\delta_\Mb\Omega^2_0\left(M;\mathbb{K}\right)$, for any given $\psi\in\Loc_\copyright\left(\Mb,\Nb\right)$, $\Mb,\Nb\in\Loc_\copyright$.
For example, $\mathcal{F}\psi$ is well-defined because the
push-forward of (co)exact forms is obviously (co)exact;  $\mathcal{F}\psi$ is also symplectic as a result of the diffeomorphism invariance of integration and $\psi^*G_\textbf{N}\psi_*=G_\textbf{M}$ (cf. \cite[Sec.3]{FeV12}). In this way, we gain two functors
\begin{align*}
\mathcal{F}:\Loc_\copyright\longrightarrow\SymplK&&\text{and}&&
\mathcal{A}:\Loc_\copyright\longrightarrow\SymplK.
\end{align*}
It is straightforward to see that $\Omega^2_0\left(M;\mathbb{K}\right)\ni\omega\mapsto\delta_\Mb\omega\in\delta_\Mb\Omega^2_0\left(M;\mathbb{K}\right)$ gives rise to a $\SymplK$-isomorphism $\eta_\Mb:\mathcal{F}\Mb\to\mathcal{A}\Mb$ for each $\Mb\in\Loc_\copyright$ and that the family $\{\eta_\Mb\}_{\Mb\in\Loc_\copyright}$ thus obtained form the components of a natural isomorphism $\eta:\mathcal{F}\dot{\rightarrow}\mathcal{A}$. Thus, $\mathcal{F}$ and $\mathcal{A}$ are naturally isomorphic (on $\Loc_\copyright$), i.e. equivalent physical theories.

The functor $\mathcal{F}$ also admits automorphisms corresponding to electromagnetic duality. To be specific, in each
$\Mb$, the Hodge-$*$ is a linear isomorphism of $\Omega_0^2(\Mb)$ to itself. As $\delta_\Mb*_\Mb = (-1)^{p+1}*_\Mb d_\Mb$  and $d_\Mb *_\Mb = (-1)^p *_\Mb \delta_\Mb$ on $\Omega^p(\Mb)$,
it is easily seen that $*_\Mb$ induces an isomorphism of the quotient
space $\left[\Omega^2_0\left(M;\mathbb{K}\right)\right]$ given
by $[\omega]\mapsto [*_\Mb\omega]$, and evidently obeys
$\mathcal{F}\psi[*_\Mb\omega]= [*_\Nb \psi_* \omega]$
for every morphism $\psi:\Mb\to\Nb$ in $\Loc_\copyright$. 
At the level of solutions, $d_\Mb G_\Mb\delta_\Mb *_\Mb\omega=-*_\Mb d_\Mb G_\Mb\delta_\Mb\omega$ for $\omega\in\Omega_0^2(\Mb)$ and one easily derives from this that
\begin{flalign*}
\textswab{w}_\Mb\left(\left[*_\Mb\omega\right],\left[*_\Mb\eta\right]\right)&=
-\int_M d_\Mb G_{\Mb}\delta_\Mb*_\Mb\omega\wedge*_\Mb *_\Mb\eta =
-\int_M d_\Mb G_{\Mb}\delta_\Mb\omega\wedge*_\Mb\eta \\
&=
\textswab{w}_\Mb\left(\left[\omega\right],\left[\eta\right]\right),&&\makebox[0pt][r]{$\left[\omega\right],\left[\eta\right]\in\left[\Omega^2_0\left(M;\mathbb{K}\right)\right]$.}
\end{flalign*}
From these results, it follows that the electromagnetic duality rotations 
\begin{flalign*}
&&\Theta_\Mb(\alpha)[\omega]= [\cos\alpha\, \omega + \sin\alpha *_\Mb\omega]
&&\omega\in\Omega_0^2(\Mb),~\Mb\in\Loc_\copyright
\end{flalign*}
yield automorphisms $\Theta(\alpha)\in\Aut(\mathcal{F})$ for $\alpha\in\mathbb{R}$;
as $\Theta(\alpha)\Theta(\beta)=\Theta(\alpha+\beta)$ and $\Theta(\alpha+2\pi)=\Theta(\alpha)$
for all $\alpha,\beta\in\mathbb{R}$,
we see that there is a faithful homomorphism from $U(1)$ into $\Aut(\mathcal{F})$. 

In \cite{Few13} the automorphisms of a locally covariant theory have been identified
as its global gauge transformations. This raises an interesting question, because the electromagnetic
duality is not a symmetry of the Maxwell Lagrangian $\mathscr{L}=-\frac{1}{4} F\wedge *F$, which changes sign under $F\mapsto *F$
and one might be concerned that the presence of these automorphisms is an indication that the theory $\mathcal{F}$ is 
not a true reflection of the original physics. Against this, we note that Maxwell Lagrangian has other unusual
properties: in particular, the field equations obtained by variation with respect to $F$ are trivial. The Maxwell
equations can be derived from the Lagrangian, however, by demanding conservation of the stress-energy
tensor constructed by varying the action with respect to the metric. As electromagnetic duality
rotations leave the stress-energy tensor 
invariant, there is good reason to accept them as symmetries of the theory.

\subsection{Extensions to non-contractible spacetimes}

In the previous subsection we obtained a satisfactory description
of the free Maxwell theory on contractible spacetimes. At various
stages in the discussion, we made use of contractibility to identify
various spaces of (co)closed forms as being (co)exact. The extension
of Maxwell theory (in the above form) to non-contractible spacetimes  presents various choices,
because electromagnetism is sensitive to the topology of the underlying spacetime. Indeed, a non-trivial spacetime topology is used for mathematical discussions related to the Aharonov-Bohm effect (cf.\ e.g.\ \cite{DHS12}) but there are also investigations of the physical relevance of a non-trivial spacetime topology purely in terms of the field strength tensor. For example, \cite{AS80} discussed the quantised free Maxwell field in terms of the field strength tensor on the Schwarzschild-Kruskal spacetime, which has the topology of $\mathbb{R}\times\mathbb{R}\times S^2$. Not every field strength tensor $F$ can be derived from a vector potential $A$ via the relation $F=dA$ on the Schwarzschild-Kruskal spacetime, ultimately leading to a two-parameter family of unitarily inequivalent representations of the canonical commutation relations labelled by topological (electric and magnetic) charges. This feature is characteristic of spacetimes with non-vanishing second de Rham cohomology group and ultimately prevents one from having classical and quantised theories in the usual, straightforward manner.

In order to deal with non-trivial topologies and to analyse the impact they have on the theory, Fredenhagen has suggested the use of (an analogue of) the universal algebra construction~\cite{Fre90,FRS92,Fre93}, to obtain the minimal description compatible with, and unifying, the local descriptions of the theory on contractible subregions of the spacetime. This was addressed in \cite[Appx A]{Hol08} and carried out in detail in \cite{DL12}.   
A similar construction can be carried out at the classical level and results in a  `universal' (complexified if $\mathbb{K}=\mathbb{C}$) pre-symplectic
space. The resulting model can be given concretely as follows. 
For each $\Mb\in\Loc$, 
$\mathcal{F}_u$ is defined as the (complexified if $\mathbb{K}=\mathbb{C}$) pre-symplectic space  $\mathcal{F}_u\Mb:=\left(\left[\Omega^2_0\left(M;\mathbb{K}\right)\right],\textswab{w}_{u\Mb},\overline{\phantom{n}}\right)$, where (omitting 
the complex conjugation if $\mathbb{K}=\mathbb{R}$)
\begin{align}\label{Fu-classic}
\begin{aligned}
\left[\Omega^2_0\left(M;\mathbb{K}\right)\right]:=\Omega^2_0\left(M;\mathbb{K}\right)\big/\left(d_\Mb\Omega^1_0\left(M;\mathbb{K}\right)\oplus\delta_\Mb\Omega^3_0\left(M;\mathbb{K}\right)\right),&&\\
\textswab{w}_{u\Mb}\left(\left[\omega\right],\left[\eta\right]\right):=-\int_MG_{\Mb}\delta_\Mb\omega\wedge*_\Mb\delta_\Mb\eta,\quad \overline{\left[\omega\right]}:=\left[\overline{\omega}\right],&&
\left[\omega\right],\left[\eta\right]\in\left[\Omega^2_0\left(M;\mathbb{K}\right)\right],
\end{aligned}
\end{align}
which is well-defined as a consequence of Lemma~\ref{lem ur sym}. 
On contractible spacetimes $\mathcal{F}_u\Mb$ coincides precisely with $\mathcal{F}\Mb$ defined by \eqref{F-classic}. 
However, the bilinear form $\textswab{w}_{u\Mb}$ is degenerate
on spacetimes with non-trivial $H^2_{dR}(M)$ (cf.\ \cite[Prop.3.3]{DL12}); indeed, Lemma~\ref{lem ur sym} entails
a linear isomorphism
\begin{flalign*}
H^2_{dR,c}(M)\oplus H^2_{dR,c}(M) &\longrightarrow \rad \textswab{w}_{u\Mb} \\
[\alpha]\oplus[\beta] &\longmapsto [\alpha+ *_\Mb \beta]
\end{flalign*} 
where the square brackets on the left are cohomology classes. 
Elements in $\rad \textswab{w}_{u\Mb}$ representing
a class in $H^2_{dR,c}(M)$ will be called {\em magnetic topological degeneracies}; elements whose Hodge dual represents a class in
$H^2_{dR,c}(M)$ will be called {\em electric topological degeneracies}.
 On any morphism
$\psi:\Mb\to\Nb$ in $\Loc$, we set $\mathcal{F}_u\psi [\omega]=
[\psi_*\omega]$, again extending the definition of $\mathcal{F}$.
The linear map $\mathcal{F}_u\psi$ is well-defined and preserves the (complexified if $\mathbb{K}=\mathbb{C}$) pre-symplectic forms for the same reasons as in contractible spacetimes;
it is clear that we have a functor $\mathcal{F}_u:\Loc\to\pSymplK$. 
It is important to note that the morphism $\mathcal{F}_u\psi$ need
not be injective; the extreme case is where $H^2_{dR}(M)\neq 0$ but $H_{dR}^2(N)=0$, in which case $\ker \mathcal{F}_u\psi =
\rad \textswab{w}_{u\Mb}$. 

Instead of quotienting by the direct sum of exact and coexact forms, we may form quotients by the larger direct sum of closed and coclosed forms, thus obtaining a {\em reduced theory}: on each $\Mb\in\Loc$, $\widetilde{\mathcal{F}}\Mb:=\left(\L\Omega^2_0\left(M;\mathbb{K}\right)\R, \tilde{\textswab{w}}_\Mb,\overline{\phantom{n}}\right)$, where
\begin{align}\label{H-classic}
\begin{aligned}
\L\Omega^2_0\left(M;\mathbb{K}\right)\R:=\Omega^2_0\left(M;\mathbb{K}\right)\big/\left(\Omega^2_{0,d}\left(M;\mathbb{K}\right)\oplus\Omega^2_{0,\delta}\left(M;\mathbb{K}\right)\right),&&\\
\tilde{\textswab{w}}_\Mb\left(\omegaa,\etaa\right):=-\int_MG_{\Mb}\delta_\Mb\omega\wedge*_\Mb\delta_\Mb\eta,\quad \overline{\omegaa}:=\L\overline{\omega}\R,&&
\omegaa,\etaa\in\L\Omega^2_0\left(M;\mathbb{K}\right)\R,
\end{aligned}
\end{align}
(omitting complex conjugation if $\mathbb{K}=\mathbb{R}$), is a $\SymplK$-object as a consequence
of  Lemma~\ref{lem ur sym}. 
Since the pushforward of compactly supported $\mathbb{K}$-valued differential forms intertwines with the exterior derivative and the exterior coderivative for each $\Loc$-morphism, 
\begin{align*}
\widetilde{\mathcal{F}}\psi:\widetilde{\mathcal{F}}\Mb\longrightarrow\widetilde{\mathcal{F}}\Nb,\enspace\omegaa\longmapsto\L\psi_*\omega\R
\end{align*}
is a well-defined $\SymplK$-morphism for $\psi\in\Loc\left(\Mb,\Nb\right)$ and $\Mb,\Nb\in\Loc$. In this way, the reduced theory  $\widetilde{\mathcal{F}}:\Loc\to\SymplK$ is a classical LCT. 

If the second de Rham cohomology group $H^2_{dR}\left(M;\mathbb{K}\right)$ of $\Mb\in\Loc$ vanishes, which implies $H^2_{dR,c}\left(M;\mathbb{K}\right)=0$ by Poincar\'{e} duality, we find $\Omega^2_{0,d}\left(M;\mathbb{K}\right)=d_\Mb\Omega^1_0\left(M;\mathbb{K}\right)$ and $\Omega^2_{0,\delta}\left(M;\mathbb{K}\right)=\delta_\Mb\Omega^3_0\left(M;\mathbb{K}\right)$. Thus, $\L\Omega^2_0\left(M;\mathbb{K}\right)\R=\left[\Omega^2_0\left(M;\mathbb{K}\right)\right]$ and $\widetilde{\mathcal{F}}\Mb=\mathcal{F}_u\Mb$ for each such $\Loc$-object; in particular,
this holds for all objects of $\Loc_\copyright$. More generally, it is clear that $\widetilde{\mathcal{F}}\Mb$ is precisely obtained from $\mathcal{F}_u\Mb$ by quotienting out its radical. 
 
Our reduced theory is closely
related to the ``charge-zero phase space functor'' for electromagnetism given in \cite[\S 7]{BDS13}. The latter functor actually yields degenerate
presymplectic spaces; however, as pointed out in \cite{FeSch14}, the
the treatment of affine theories used in \cite{BDS13} should be corrected; once this is done their approach would coincide with our reduced theory. 

Of course, it would have been possible to start in the A-description and then pass to the
corresponding `universal' (complexified if $\mathbb{K}=\mathbb{C}$) pre-symplectic space $\mathcal{A}_u:\Loc\to\pSymplK$ obtained from
$\mathcal{A}:\Loc_\copyright\to\pSymplK$ in the same way as $\mathcal{F}_u$. As $\mathcal{A}$ and $\mathcal{F}$ are
naturally isomorphic theories, however, it follows on abstract grounds that $\mathcal{A}_u$ is
naturally isomorphic to $\mathcal{F}_u$. Explicitly, $\mathcal{A}_u\Mb$ is given simply
by the formulae in \eqref{A-classic} but with $\Mb\in\Loc$ allowed to be arbitrary,
and there is a natural isomorphism $\eta_u:\mathcal{F}_u\dot{\rightarrow}\mathcal{A}_u\Mb$
with components $\eta_{u\,\Mb}[\omega]=[\delta_\Mb\omega]$. 

Finally, we note that the automorphisms $\Theta(\alpha)\in\Aut(\mathcal{F})$ implementing electromagnetic duality rotations lift to automorphisms of both the universal and reduced theories.


\subsection{Quantisation}\label{subsec contractible field algebra}

The models described above can be conveniently quantised as infinitesimal Weyl algebras
by means of a quantisation functor $\mathcal{Q}:\pSymplC\to\uAlg$, which is defined as follows. 
For any $\left(V,\omega,C\right)\in\pSymplC$, let $\mathcal{Q}\left(V,\omega,C\right)$ be
the complex symmetric tensor vector space 
\begin{flalign*}
\mathcal{Q}\left(V,\omega,C\right) &=\Gamma_\odot(V) \stackrel{\text{def}}{=} \bigoplus_{n\in\mathbb{N}_0} V^{\odot n} ,
\end{flalign*}
equipped with the product uniquely determined by the requirements
\begin{flalign*} 
u^{\odot m} \cdot v^{\odot n} = \sum_{r=0}^{\min\{m,n\}}
\left(\frac{i\omega(u,v)}{2}\right)^r \frac{m!n!}{r!(m-r)!(n-r)!}
S\left( u^{\otimes (m-r)}\otimes v^{\otimes (n-r)}\right),
\end{flalign*}
for all $m,n\in\mathbb{N}_0$ and $u,v\in V$. Here $S$ denotes symmetrisation,
and we use the convention $u^{\odot}=1\in V^{\odot 0}=\mathbb{C}$. 
The $*$-operation is uniquely defined by $(u^{\odot n})^* = (Cu)^{\odot n}$. 
To any morphism $f\in\pSymplC\left(\left(V,\omega,C\right),\left(V',\omega',C'\right)\right)$, 
we assign the unital $*$-homomorphism $(\mathcal{Q}f)=\Gamma_\odot(f)
\in\uAlg\left(\mathcal{Q}\left(V,\omega,C\right),(\mathcal{Q}\left(V',\omega',C'\right)\right)$.
A discussion of a related functor can be found, for example, in \cite[\S 5]{FeV12}.
In fact, that reference concerns the restriction of $\mathcal{Q}$ to
$\mpSymplC$ which takes its values in $\umAlg$ -- here, this functor will
be denoted $\mathcal{Q}^m$. The proof that
our $\mathcal{Q}$ is indeed a functor is simply obtained by dropping any
references to injectivity in the proof of \cite[Prop.5.1]{FeV12}, a result
which also shows that non-degeneracy of $\left(V,\omega,C\right)$ 
implies that $\mathcal{Q}\left(V,\omega,C\right)$ is simple (see also \cite[Scholium 7.1]{BSZ92}).

Applying $\mathcal{Q}$ and $\mathcal{Q}^m$, we obtain 
quantised theories 
\begin{align*}
\textswab{F}:=\mathcal{Q}^m\circ\mathcal{F}:\Loc_\copyright\longrightarrow\umAlg&&\text{and}&&
\textswab{A}:=\mathcal{Q}^m\circ\mathcal{A}:\Loc_\copyright\longrightarrow\umAlg
\end{align*}
on contractible spacetimes
and also quantisations of the universal F- and A-theories
\begin{align*}
\textswab{F}_u:=\mathcal{Q}\circ\mathcal{F}_u:\Loc\longrightarrow\uAlg &&\text{and}
&&\textswab{A}_u:=\mathcal{Q}\circ\mathcal{A}_u:\Loc\longrightarrow\uAlg,
\end{align*}
and also of the reduced theory $\widetilde{\textswab{F}}:=\mathcal{Q}^m\circ\widetilde{\mathcal{F}}:\Loc\longrightarrow\umAlg$. It is clear that $\widetilde{\textswab{F}}\Mb = \textswab{F}_u\Mb$
if $M$ has trivial second de Rham cohomology, and that these algebras coincide with $\textswab{F}\Mb$
if $M$ is contractible. Moreover, the algebras $\textswab{F}\Mb$, $\textswab{A}\Mb$ and
$\widetilde{\textswab{F}}\Mb$ are  simple for all $\Mb\in\Loc$, while $\textswab{F}_u\Mb$ and $\textswab{A}_u\Mb$  have non-trivial centres if $H^2_{dR}(M)\neq 0$ (see \cite[\S 3.6]{DL12}). 

These theories can also be described in terms of fields. For example, let $\Mb\in\Loc$ and define, for
each $\omega\in\Omega_0^2(M;\mathbb{C})$, the element $\widehat{\textbf{F}}_\Mb\left(\omega\right)=(0,\omega,\ldots)\in\Gamma_\odot(\mathcal{F}_u\Mb)$.  Then one readily sees that the $\widehat{\textbf{F}}_\Mb\left(\omega\right)$ generate
$\textswab{F}_u\Mb$ and obey the following relations (cf.\ \cite[Def.3.1]{DL12}): 
\begin{itemize}\itemsep -0.75em 
\item Linearity and hermiticity: 
\begin{flalign*}
&&\widehat{\textbf{F}}_\Mb\left(\lambda\omega+\mu\eta\right)=
\lambda\widehat{\textbf{F}}_\Mb\left(\omega\right)+\mu\widehat{\textbf{F}}_\Mb\left(\eta\right)
&&\text{and}&& \widehat{\textbf{F}}_\Mb\left(\omega\right)^*=\widehat{\textbf{F}}_\Mb\left(\overline{\omega}\right) &&\\
&&&&&&&&\makebox[0pt][r]{$\forall\lambda,\mu\in\mathbb{C},\,\forall\omega,\eta\in\Omega^2_0\left(M;\mathbb{C}\right)$.}
\end{flalign*}
\item  Free Maxwell equations in the weak sense:  
\begin{flalign*}
&&\widehat{\textbf{F}}_\Mb\left(d_\Mb\theta\right)=0&&\text{and}&&
\widehat{\textbf{F}}_\Mb\left(\delta_\Mb\eta\right)=0&&
\forall\theta\in\Omega^1_0\left(M;\mathbb{K}\right),\,\forall\eta\in\Omega^3_0\left(M;\mathbb{C}\right).
\end{flalign*}
\item Commutation relations:\footnote{Also known as Lichnerowicz's commutation relations -- see the remark in  \cite[\S 4]{Dim92} and \cite{Lich61}.}
\begin{flalign*}
&&\left[\widehat{\textbf{F}}_\Mb\left(\omega\right),\widehat{\textbf{F}}_\Mb\left(\eta\right)\right]=
\left(-\iu\int_MG_\Mb\delta_\Mb\omega\wedge*_\Mb\delta_\Mb\eta\right)\cdot1_{\textswab{F}_u\Mb}&&\forall\omega,\eta\in\Omega^2_0\left(M;\mathbb{C}\right).
\end{flalign*}
\end{itemize}
Under morphisms $\psi:\Mb\to\Nb$, we have 
$(\textswab{F}_u\psi)(\widehat{\textbf{F}}_\Mb\left(\omega\right))= \widehat{\textbf{F}}_\Nb\left(\psi_*\omega\right)$ which shows that
 the $\widehat{\textbf{F}}_\Mb$ constitute a locally covariant field in the sense of \cite{BFV03}.
One may define fields for the theory $\widetilde{\textswab{F}}$ in a similar way, with the 
difference that 
the axiom for Maxwell's equations is now replaced by $\widetilde{\textbf{F}}_\Mb\left(\omega\right)=0$
for all $\omega\in\Omega^2_{0,d}\left(M;\mathbb{C}\right)\oplus\Omega^2_{0,\delta}\left(M;\mathbb{C}\right)$.

In the A-description, we define $\widehat{\textbf{[A]}}_\Mb\left(\theta\right)=(0,\theta,\ldots)\in\Gamma_\odot(\mathcal{F}_u\Mb)$ ($\theta\in\delta_\Mb\Omega^2_0\left(M;\mathbb{C}\right)$) thereby obtaining generators obeying the following relations \cite{Dim92,FePf03,Pf09,DHS12}:
\begin{itemize}
\itemsep-0.75em 
\item[$\bullet$] Linearity and hermiticity: 
\begin{flalign*}
&&\widehat{\textbf{[A]}}_\Mb\left(\lambda\theta+\mu\phi\right)=
\lambda\widehat{\textbf{[A]}}_\Mb\left(\theta\right)+\mu\widehat{\textbf{[A]}}_\Mb\left(\phi\right)&&
\text{and}&&
\widehat{\textbf{[A]}}_\Mb\left(\theta\right)^*=\widehat{\textbf{[A]}}_\Mb\left(\overline{\theta}\right)&&
\\ 
&&&&&&&&\makebox[0pt][r]{$\forall\lambda,\mu\in\mathbb{C},\, 
\forall\theta,\phi\in\delta_\Mb\Omega^2_0\left(M;\mathbb{C}\right)$.}
\end{flalign*}
\item[$\bullet$] Free Maxwell equations in the weak sense:
\begin{flalign*}
&&\widehat{\textbf{[A]}}_\Mb\left(\delta_\Mb d_\Mb\theta\right)=0&&\forall\theta\in\Omega^1_0\left(M;\mathbb{C}\right).
\end{flalign*}
\item[$\bullet$] Commutation relations:
\begin{flalign*}
&&\left[\widehat{\textbf{[A]}}_\Mb\left(\theta\right),\widehat{\textbf{[A]}}_\Mb\left(\phi\right)\right]=
\left(-\iu\int_MG_\Mb\theta\wedge*_\Mb\phi\right)\cdot1_{\textswab{A}_u\Mb}&&\forall\theta,\phi\in\delta_\Mb\Omega^2_0\left(M;\mathbb{C}\right).
\end{flalign*}
\end{itemize}
Under a morphism $\psi:\Mb\to\Nb$, we have $\left(\textswab{A}_u\psi\right)\left(\widehat{\textbf{[A]}}_\Mb\left(\theta\right)\right)=\widehat{\textbf{[A]}}_\Nb\left(\psi_*\theta\right)$ for all $\theta\in\delta_\Mb\Omega^2_0\left(M;\mathbb{C}\right)$. 

The theories $\textswab{F}$ and $\textswab{A}$ are actually equivalent because
 the natural isomorphism $\eta_u:\mathcal{F}_u\dot{\rightarrow}\mathcal{A}_u$  lifts to a natural isomorphism $\mathcal{Q}\star\eta_u:\textswab{F}_u\dot{\rightarrow}\textswab{A}_u$, $\left(\mathcal{Q}\star\eta_u\right)_\Mb:=\mathcal{Q}\left(\eta_{u\,\Mb}\right)$ for $\Mb\in\Loc$.
This precisely generalises the ``natural algebraic relation'' between the Borchers-Uhlmann algebras for the F- and the A-descriptions discussed in \cite{Bong77} for Minkowski space. Explicitly,
$\eta_{u\,\Mb}\widehat{\textbf{F}}_\Mb\left(\omega\right)=
\widehat{\textbf{[A]}}_\Mb\left(\delta_\Mb\omega\right)$, $\omega\in\Omega^2_0\left(M;\mathbb{C}\right)$, which is the weak analogue of the familiar relation $F=dA$. 
Owing to this equivalence, all statements about the classical and the quantised universal free F-theory 
apply equally to the classical and the quantised universal free A-theory. Choosing the classical universal free $F$-theory over the A-theory and vice versa has no physical significance and purely expresses a different point of view on the same theory. In the following sections, we we take the point of view of the F-description, which slightly simplifies some arguments. One may
introduce classical and quantised reduced A-theories isomorphic to the corresponding reduced F-theories. 

Finally, the electromagnetic duality rotation automorphisms possessed by all the classical theories lift immediately to the quantised theories by the action of the quantisation functor.   

\section{Dynamical locality of the universal theory}\label{sec dynamical locality universal theory}

\subsection{The universal theory fails local covariance}\label{sec:univ fails}
It was already pointed out in \cite[\S 3.7]{DL12} that the quantised universal free F-theory $\textswab{F}_u:\Loc\to\uAlg$ is not a LCQFT according to \cite{BFV03} because morphisms corresponding to spacetime embeddings are not always injective. The same is true for the classical universal free F-theory $\mathcal{F}_u:\Loc\to\pSymplK$. Indeed, consider any $\Loc$-morphism $\psi:\Mb\to\Nb$  between objects
obeying $H^2_{dR}\left(M;\mathbb{K}\right)\neq0$ and   $H^2_{dR}\left(N;\mathbb{K}\right)=0$ (for example, let $\Nb$ be Minkowski spacetime, $\Mb$ the Cauchy development of $\left\{0\right\}\times\left\{\left(x,y,z\right)^\top\in\mathbb{R}^3\mid x^2+y^2+z^2>1\right\}$ in $\Nb$, 
which has $H^2_{dR}\left(M;\mathbb{K}\right)\cong\mathbb{K}$, and $\psi\in\Loc\left(\Mb,\Nb\right)$ the inclusion map). 
Let $\omega\in\Omega^2_{0,d}\left(M;\mathbb{K}\right)\setminus d_\Mb\Omega^2_{0}\left(M;\mathbb{K}\right)$, which is nonempty
because  $H^2_{dR,c}\left(M;\mathbb{K}\right)\neq 0$ by Poincar\'e duality. Then $\omega$ cannot be written
in the form $\omega=d_\Mb\theta+\delta_\Mb\eta$ for $\theta\in\Omega^1_0\left(M;\mathbb{K}\right)$ and $\eta\in\Omega^3_0\left(M;\mathbb{K}\right)$,\footnote{Otherwise, 
$\delta_\Mb\eta=G^\text{ret}_\Mb\Box_\Mb\delta_\Mb\eta=
-G^\text{ret}_\Mb\delta_\Mb d_\Mb\delta_\Mb\eta=
-G^\text{ret}_\Mb\delta_\Mb d_\Mb(d_\Mb\theta-\omega)=0$, so $\omega=d_\Mb\theta$, a contradiction.}  so $\left[\omega\right]\neq0\in\left[\Omega^2_0\left(M;\mathbb{K}\right)\right]$. However, the push-forward
$\psi_*\omega\in\Omega^2_0\left(N;\mathbb{K}\right)$
obeys $d_\Nb\psi_*\omega=\psi_*d_\Mb\omega=0$ and hence
$\psi_*\omega\in d_\Nb\Omega^1_0\left(N;\mathbb{K}\right)\oplus \delta_\Nb\Omega^3_0\left(N;\mathbb{K}\right)$ because
$H^2_{dR,c}\left(N;\mathbb{K}\right)=0$ by Poincar\'e duality. 
Thus
$\left(\mathcal{F}\!_u\psi\right)\left[\omega\right]
=\left[\psi_*\omega\right]=0\in\left[\Omega^2_0\left(N;\mathbb{K}\right)\right]$ and 
$\left(\textswab{F}_u\psi\right)
\left(\widehat{\textbf{F}}_\Mb\left(\omega\right)\right) =\widehat{\textbf{F}}_\Nb\left(\psi_*\omega\right) =0_{\textswab{F}_u\Nb}$, so neither
$\mathcal{F}\!_u\psi$ nor $\textswab{F}_u\psi$ is injective. 

A similar argument applies to $\omega\in\Omega^2_{0,\delta}\left(M;\mathbb{K}\right)\setminus \delta_\Mb\Omega^2_{0}\left(M;\mathbb{K}\right)$. The elements just described in this and the last paragraph are precisely the ones that span the radical of $\textswab{w}_{u\Mb}$ and the centre of $\textswab{F}_u\Mb$, respectively, $\Mb\in\Loc$ (cf. \cite[Prop.3.3]{DL12}). Hence, local covariance of $\mathcal{F}\!_u$ and $\textswab{F}_u$ is precisely spoiled by the radical elements and the central elements, respectively.

However, $\textswab{F}_u$ is still a causal functor, owing to the form of Lichnerowicz's commutator, and as we will see shortly, both $\mathcal{F}\!_u$ and $\textswab{F}_u$ obey the time-slice axiom, i.e. $\mathcal{F}\!_u\psi$ is a $\pSymplK$-isomorphism and $\textswab{F}_u\psi$ is an $\uAlg$-isomorphism whenever $\psi\in\Loc\left(\Mb,\Nb\right)$ is Cauchy.

\subsection{The universal theory obeys the time-slice axiom}\label{subsec time-slice axiom universal F-theory}
We start with some helpful, more general statements, which will allow us to show the validity of the time-slice axiom and to compute inverses. For the rest of this subsection, let $\psi\in\Loc\left(\Mb,\Nb\right)$ be Cauchy, $\xi=(E,N,\pi,V)$ a smooth $\mathbb{K}$-vector bundle over $N$ and $P:\Gamma^\infty(\xi)\to\Gamma^\infty(\xi)$ a normally hyperbolic differential operator of metric type. 

\begin{defn}\label{def time-slice map} A \emph{time-slice map} for $\left(\psi, \xi, P\right)$ is a $\mathbb{K}$-linear map $L:\Gamma_0^\infty\left(\xi\right)\to\Gamma_0^\infty\left(\xi\right)$ satisfying 
\[
\left(\id_{\Gamma_0^\infty\left(\xi\right)}-PL\right)\Gamma_0^\infty\left(\xi\right)\subseteq\mathfrak{i}_{\xi|_{\psi\left(M\right)}*}\Gamma_0^\infty\left(\xi|_{\psi\left(M\right)}\right),
\]
where $\mathfrak{i}_{\xi|_{\psi\left(M\right)}*}:\Gamma_0^\infty\left(\xi|_{\psi\left(M\right)}\right)\to\Gamma^\infty_0\left(\xi\right)$ denotes the pushforward of compactly supported smooth cross-sections in the restricted smooth $\mathbb{K}$-vector bundle of $\xi$ to $\psi\left(M\right)$ along the bundle inclusion $\mathfrak{i}_{\xi|_{\psi\left(M\right)}}:\xi|_{\psi\left(M\right)}\to\xi$.
\end{defn}

If a particular time-slice map is understood, we will write
\[
\sigma=\sigma_{\Euro}+P\sigma_{\!\pounds}
\]
for the corresponding decomposition $\sigma_{\!\pounds}:=L\sigma$, $\sigma_{\Euro}:=\sigma-P\sigma_{\!\pounds}$.

Time-slice maps exist by slight modification of a standard construction: fix any two smooth spacelike Cauchy surfaces $\Sigma^f$ and $\Sigma_p$ for $\Nb$ such that $\Sigma^f,\Sigma_p\subseteq\psi\left(M\right)$ and $\Sigma^f$ lies strictly in the future of $\Sigma_p$. This can be achieved using \cite[Lem.A.2]{SPASs12} and the splitting theorem of Bernal and S\'{a}nchez \cite[Prop.2.4]{BeSa05}. Further, let $\left\{\chi^+,\chi^-\right\}$ be a smooth partition of unity subordinated to the open cover $\left\{I_\Nb^+\left(\Sigma_p\right), I_\Nb^-\left(\Sigma^f\right)\right\}$ of $N$. Define for each $\sigma\in\Gamma^\infty_0\left(\xi\right)$
\begin{align}\label{eq:tslice}
\sigma_{\!\Euro}:=\sigma-P\chi^+G^\text{adv}\sigma-P\chi^-G^\text{ret}\sigma,
\end{align}
where $G^\text{adv}$ and $G^\text{ret}$ are the advanced and the retarded Green's operator for $P$, which exist and are unique  \cite[Cor.3.4.3]{BGP07}. By the properties of $\chi^{\pm}$ and $G^\text{ret/adv}$, $\supp\sigma_{\!\Euro}$ is compactly supported in $\psi\left(M\right)$. Finally, $\sigma_{\!\pounds}\in\Gamma^\infty_0\left(\xi\right)$ is defined by $\sigma_{\!\pounds}:=\chi^+G^\text{adv}\sigma+\chi^-G^\text{ret}\sigma$.
However, many properties of time-slice maps can be proved without using a specific formula. The main technical point is that any compactly supported solution $\phi$ to the inhomogeneous equation $P\phi=\sigma$, where
$\sigma\in\Gamma_0^\infty\left(\xi\right)$, must be supported in the intersection $J_\Nb^+\left(\supp\sigma\right)\cap J_\Nb^-\left(\supp\sigma\right)$ because $\phi=G^{\text{ret}/\text{adv}}\sigma$. Let us observe

\begin{lem}\label{lem time-slice axiom 1} If $L$ is any time-slice map for $(\psi, \xi, P)$, we have
\[
L\left(\mathfrak{i}_{\xi|_{\psi\left(M\right)}*}\Gamma_0^\infty\left(\xi|_{\psi\left(M\right)}\right)\right)
\subseteq\mathfrak{i}_{\xi|_{\psi\left(M\right)}*}\Gamma_0^\infty\left(\xi|_{\psi\left(M\right)}\right)
\] 
and if $K$ is another time-slice map for $\left(\psi, \xi, P\right)$, then 
\[
\left(K-L\right)\Gamma_0^\infty\left(\xi\right)\subseteq\mathfrak{i}_{\xi|_{\psi\left(M\right)}*}\Gamma_0^\infty\left(\xi|_{\psi\left(M\right)}\right).
\]
Hence,
\[
\sigma_{\!\Euro_K}-\sigma_{\!\Euro_L}\in P\mathfrak{i}_{\xi|_{\psi\left(M\right)}*}\Gamma_0^\infty\left(\xi|_{\psi\left(M\right)}\right) ;
\]
moreover,
\[
L\sigma|_{N\setminus J_\Nb^{-/+}(\psi(M))} = G_\Nb^{\text{adv/ret}}\sigma|_{N\setminus J_\Nb^{-/+}(\psi(M))}.
\]
\end{lem}
\noindent\textbf{\textit{Proof:}} Taking any $\sigma\in\mathfrak{i}_{\xi|_{\psi\left(M\right)}*}\Gamma_0^\infty\left(\xi|_{\psi\left(M\right)}\right)$, $PL\sigma=\sigma-\left(\id_{\Gamma^\infty_0\left(\xi\right)}-PL\right)\sigma$ is (compactly) supported in $\psi\left(M\right)$. As $L\sigma$ is compactly supported, it follows that $L\sigma$ is supported in $J_\Nb^+\left(\psi\left(M\right)\right)\cap J_\Nb^-\left(\psi\left(M\right)\right)=\psi\left(M\right)$ as required. 
Next, let $\sigma\in \Gamma_0^\infty\left(\xi\right)$. Then by definition of time-slice maps, $P\left(K-L\right)\sigma$ has support
in $\psi\left(M\right)$, while $\left(K-L\right)\sigma$ has compact support. Thus
$\left(K-L\right)\sigma$ is (compactly) supported in $J_\Nb^+\left(\psi\left(M\right)\right)\cap J_\Nb^-\left(\psi\left(M\right)\right)=\psi\left(M\right)$. 
The penultimate formula follows from this and the definition $\sigma_{\Euro}:=\sigma-P\sigma_{\!\pounds}=\sigma-PL\sigma$ for $\sigma\in \Gamma_0^\infty\left(\xi\right)$. Finally, our result shows that the action of any timeslice
map on $\sigma$ is fixed modulo terms compactly supported in $\psi(M)$. Outside this set, 
all timeslice maps agree, so we may use the formula implicit in \eqref{eq:tslice} to obtain
the final result. \hfill\SquareCastShadowTopRight\par\bigskip

As a digression, the existence of a time-slice map for $\left(\psi,\xi, P\right)$ implies that the following is a short exact sequence of $\mathbb{K}$-linear maps
\[
0\longrightarrow P\left(\mathfrak{i}_{\xi|_{\psi\left(M\right)}*}\Gamma_0^\infty\left(\xi|_{\psi\left(M\right)}\right)\right)
\xlongrightarrow{\alpha}\Gamma_0^\infty\left(\xi\right)\oplus P\left(\mathfrak{i}_{\xi|_{\psi\left(M\right)}*}\Gamma_0^\infty\left(\xi|_{\psi\left(M\right)}\right)\right)
\xlongrightarrow{\beta}\Gamma_0^\infty\left(\xi\right)\longrightarrow 0
\]
where $\alpha:\sigma\longmapsto\left(\sigma,-\sigma\right)$ and
$\beta:\left(\sigma,\tau\right)\longmapsto\sigma+\tau$. Exactness at $P\left(\mathfrak{i}_{\xi|_{\psi\left(M\right)}*}\Gamma_0^\infty\left(\xi|_{\psi\left(M\right)}\right)\right)$ is immediate because $\alpha$ is injective; moreover its image is precisely the kernel of $\beta$, so we have exactness at $\Gamma_0^\infty\left(\xi\right)\oplus P\left(\mathfrak{i}_{\xi|_{\psi\left(M\right)}*}\Gamma_0^\infty\left(\xi|_{\psi\left(M\right)}\right)\right)$. Any time-slice map $L$ for $\left(\psi,\xi,P\right)$ induces $\gamma:\Gamma_0^\infty\left(\xi\right)\to\Gamma_0^\infty\left(\xi\right)\oplus P\left(\mathfrak{i}_{\xi|_{\psi\left(M\right)}*}\Gamma_0^\infty\left(\xi|_{\psi\left(M\right)}\right)\right)$ by $\gamma:\sigma\longmapsto
\left(\sigma-PL\sigma,PL\sigma\right)$, and as $\beta\circ\gamma=\id_{\Gamma^\infty_0\left(\xi\right)}$,
it is clear that $\beta$ is surjective and we have a split short exact sequence.

\begin{lem}\label{lem time-slice axiom 2} Let $\eta=\left(D,N,\varrho,W\right)$ be a smooth $\mathbb{K}$-vector bundle with a normally hyperbolic differential operator $Q:\Gamma^\infty\left(\eta\right)\to\Gamma^\infty\left(\eta\right)$ such that $P$ and $Q$ are intertwined by a (partial) differential operator $\partial:\Gamma^\infty\left(\xi\right)\to\Gamma^\infty\left(\eta\right)$, i.e. $\partial\circ P=Q\circ\partial$. Suppose $L$ and $K$ are time-slice maps for $\left(\psi,\xi, P\right)$ and $\left(\psi, \eta, Q\right)$, then for any $\sigma\in\Gamma_0^\infty\left(\xi\right)$,
\begin{equation*}
\partial L\sigma- K\partial\sigma\in\mathfrak{i}_{\eta|_{\psi\left(M\right)}*}\Gamma_0^\infty\left(\eta\right)
\end{equation*}
and accordingly
\begin{equation*}
\left(\partial\sigma\right)_{\!\Euro_K}-\partial\sigma_{\!\Euro_L}
=Q\left(\partial L\sigma- K\partial\sigma\right)\in Q\left(\mathfrak{i}_{\eta|_{\psi\left(M\right)}*}\Gamma_0^\infty\left(\eta\right)\right).
\end{equation*}
\end{lem}
\noindent\textbf{\textit{Proof:}} We calculate for $\sigma\in\Gamma^\infty_0\left(\xi\right)$
\begin{align*}
Q\left(\partial L\sigma-K\partial\sigma\right)&=\partial PL\sigma-QK\partial\sigma=\partial\left(\sigma-\sigma_{\!\Euro_L}\right)- \left(\partial\sigma-\left(\partial\sigma\right)_{\!\Euro_K}\right)\\
&=\left(\partial\sigma\right)_{\!\Euro_K}-\partial\sigma_{\!\Euro_L}\in\mathfrak{i}_{\eta|_{\psi\left(M\right)}*}\Gamma^\infty_0\left(\eta\right).
\end{align*}
Hence, $\partial L\sigma-K\partial\sigma$ is compactly supported in $\psi\left(M\right)$ and the remaining assertion follows.\hfill\SquareCastShadowTopRight\par\bigskip

Finally, let us apply this to differential forms with a view to the description of electromagnetism. Let our smooth $\mathbb{K}$-vector bundles be the (complexified if $\mathbb{K}=\mathbb{C}$) $p$-th exterior power $\lambda^p_N$ of the cotangent bundle $\tau^*_N$ of $N$ for $p\geq0$ and let $\psi\in\Loc\left(\Mb,\textbf{N}\right)$ be Cauchy. Then taking the appropriate wave operators as the normally hyperbolic differential operators acting on differential $p$-forms, the exterior derivative and the exterior coderivative provide intertwining operators. 
The previous lemma now gives the following. 
\begin{lem} \label{lem time-slice axiom 3}
For any time-slice map $L:\Omega_0^p\left(N;\mathbb{K}\right)\to\Omega_0^p\left(N;\mathbb{K}\right)$, we have
\begin{align*}
(d_\Nb\omega)_{\Euro}-d_\Nb\omega_{\!\Euro}\in\Box_\Nb\iota_{\psi\left(M\right)*}\Omega_0^{p+1}\left(\psi\left(M\right);\mathbb{K}\right),\,\qquad
(\delta_\Nb\omega)_{\Euro}-\delta_\Nb\omega_{\Euro}\in\Box_\Nb\iota_{\psi\left(M\right)*}\Omega_0^{p-1}\left(\psi\left(M\right);\mathbb{K}\right),
\end{align*}
for $\omega\in\Omega_0^p\left(N;\mathbb{K}\right)$. Further, if $\omega\in\Omega_{0,d}^p\left(N;\mathbb{K}\right)\oplus
\Omega_{0,\delta}^{p}\left(\psi\left(N\right);\mathbb{K}\right)$, then
\begin{equation}\label{eq:omegaeuro}
\omega_{\Euro}\in\iota_{\psi\left(M\right)*}\left(\Omega_{0,d}^p\left(\psi\left(M\right);\mathbb{K}\right) \oplus
\Omega_{0,\delta}^{p}\left(\psi\left(M\right);\mathbb{K}\right)\right).
\end{equation}
\end{lem}
\noindent\textbf{\textit{Proof:}}  The first part is a direct consequence
of Lem.~\ref{lem time-slice axiom 2}. Now suppose that $d_\textbf{N}\omega=0$, then $d_\textbf{N}L\omega\in\iota_{\psi\left(M\right)*}\Omega_0^{p+1}\left(\psi\left(M\right);\mathbb{K}\right)$, where $\iota_{\psi\left(M\right)}$ denotes the pushforward of compactly supported $\mathbb{K}$-valued differential $p$-forms ($p\geq0$) along the inclusion map $\iota_{\psi\left(M\right)}:\psi\left(M\right)\rightarrow N$. Using the fact that  $\Box_\Nb=-\left(d_\Nb\delta_\Nb+\delta_\Nb d_\Nb\right)$, we have
\begin{align*}
\omega=\omega_{\Euro}+\Box_\textbf{N}L\omega&&\text{or equivalently}&&\omega+d_\Nb\delta_\Nb L\omega=\omega_{\Euro}-\delta_\Nb d_\Nb L\omega
\end{align*}
the right-hand side of which is obviously supported in $\psi\left(M\right)$. Hence, the left-hand side of the second equation must have the same support and is in the kernel of $d_\Nb$. Thus 
\eqref{eq:omegaeuro} holds for closed $\omega$, and as the same argument applies to coexact $\omega$, the result is proved. 
\hfill\SquareCastShadowTopRight\par\bigskip 

We will now apply these general statements in order to show that $\mathcal{F}\!_u$ and $\textswab{F}_u$ obey the time-slice axiom. In the proof, we will explicitly construct the inverses of $\mathcal{F}\!_u\psi$ and $\textswab{F}_u\psi$, where $\psi\in\Loc\left(\Mb,\Nb\right)$ is Cauchy, which will be helpful when computing a concrete expression for the relative Cauchy evolution for $\mathcal{F}\!_u$ and $\textswab{F}_u$. Since functors preserve isomorphisms and $\textswab{F}_u=\mathcal{Q}\circ\mathcal{F}\!_u$ (where  $\mathcal{Q}:\pSymplK\to\uAlg$ is the quantisation functor) it is enough to concentrate on the classical universal free F-theory. 

\begin{prop}\label{prop time-slice axiom 4} For $\psi\in\Loc\left(\Mb,\Nb\right)$ Cauchy, $\mathcal{F}\!_u\psi$ is a $\pSymplK$-isomorphism whose inverse is explicitly given by
\begin{align*}
&&\left(\mathcal{F}\!_u\psi\right)^{-1}:\mathcal{F}\!_u\Nb\to\mathcal{F}\!_u\Mb,&&\left[\omega\right]\longmapsto\left[\psi^*\omega_{\Euro}\right],&&
\end{align*} 
for any time-slice map of $\left(\psi, \lambda^2_N,\Box_\Nb\right)$ and any representative $\omega$ of the equivalence class $\left[\omega\right]\in\left[\Omega^2_0\left(N;\mathbb{K}\right)\right]$.
\end{prop}
\noindent\textbf{\textit{Proof:}} By Lemma \ref{lem time-slice axiom 1} and Lemma \ref{lem time-slice axiom 2}, the map $\Xi:\mathcal{F}\!_u\Nb\to\mathcal{F}\!_u\Mb$, $\left[\omega\right]\longmapsto\left[\psi^*\omega_{\Euro}\right]$, is well-defined, i.e.\ independent of the representative of $\left[\omega\right]\in\left[\Omega^2_0\left(N;\mathbb{K}\right)\right]$ and the time-slice map chosen (cf.\ the paragraph after Lemma \ref{lem time-slice axiom 2}). It is not difficult to check that $\Xi$ is $\mathbb{K}$-linear, symplectic and intertwines with the $C$-involution in the case $\mathbb{K}=\mathbb{C}$. The computations
\begin{flalign*}
&&\left(\Xi\circ\left(\mathcal{F}\!_u\psi\right)\right)\left[\omega\right]&=\Xi \left[\psi_*\omega\right]=\left[\psi^*\left(\psi_*\omega\right)_{\Euro}\right]=\left[\psi^*\psi_*\omega\right]=\left[\omega\right]
&&&&\makebox[0pt][r]{$\forall\left[\omega\right]\in\left[\Omega^2_0\left(M;\mathbb{K}\right)\right]$,}
\end{flalign*}
where we have used Lemma \ref{lem time-slice axiom 1}, and
\begin{flalign*}
&&\left(\left(\mathcal{F}\!_u\psi\right)\circ \Xi\right)\left[\omega\right]=\left(\mathcal{F}\!_u\psi\right) \left[\psi^*\omega_{\Euro}\right]=\left[\psi_*\psi^*\omega_{\Euro}\right]=\left[\omega_{\Euro}\right]=\left[\omega\right]
&&&&\makebox[0pt][r]{$\forall\left[\omega\right]\in\left[\Omega^2_0\left(N;\mathbb{K}\right)\right]$}
\end{flalign*}
show the rest.\hfill\SquareCastShadowTopRight\par

Accordingly, both $\mathcal{F}_u$ and (applying the quantisation functor)   $\textswab{F}_u$  obey the time-slice axiom. 

\subsection{The relative Cauchy evolution of the universal theory}\label{subsec relative Cauchy evolution universal F-theory}
The explicit inverse  computed in Prop.~\ref{prop time-slice axiom 4}  allows us to compute the relative Cauchy evolution for $\mathcal{F}\!_u$ and $\textswab{F}_u$ induced by $h\in H\left(\Mb\right)$. To this end, let $L^\pm:\Omega^2_0\left(M;\mathbb{K}\right)\to\Omega^2_0\left(M;\mathbb{K}\right)$ be time-slice maps for $\left(\imath^+_\Mb\left[h\right]:\Mb^+\left[h\right]\to\Mb,\lambda^2_M,\Box_\Mb\right)$
and $\left(\imath^-_\Mb\left[h\right]:\Mb^-\left[h\right]\to\Mb[h],\lambda^2_M,\Box_{\Mb[h]}\right)$ respectively
and use the symbols `$\Euro^\pm$' to correspond to $L^\pm$. Then we have, for any $\left[\omega\right]\in\left[\Omega^2_0\left(M;\mathbb{K}\right)\right]$,
\[
\rce^{\mathcal{F}\!_u}_{\Mb}\left[h\right]\left[\omega\right] = \left[(\omega_{\Euro^+})_{\Euro^-}\right]
= \left[\omega_{\Euro^+}\right] - \left[\Box_{\Mb[h]}L^-\omega_{\Euro^+}\right]
= \left[\omega \right] + \left[(\Box_\Mb -\Box_{\Mb[h]})L^-\omega_{\Euro^+}\right]
\]
where we have used the fact that $L^-\omega_{\Euro^+}$ is compactly supported and hence 
$\left[\Box_\Mb L^-\omega_{\Euro^+}\right]=0$. Now $\Box_\Mb$ and $\Box_{\Mb[h]}$ differ only on the support of $h$,
which lies outside and to the future of the range of $\imath^-_\Mb\left[h\right]$, allowing us to replace $L^-$ by $G_{\Mb[h]}^{\textrm{adv}}$ 
(by the last part of Lem.~\ref{lem time-slice axiom 1}). Hence 
\[
\rce^{\mathcal{F}\!_u}_{\Mb}\left[h\right]\left[\omega\right]  
= \left[\omega \right] + \left[(\Box_\Mb -\Box_{\Mb[h]})G_{\Mb[h]}^{\textrm{adv}} \omega_{\Euro^+}\right]
= \left[\omega \right] - \left[(\Box_\Mb -\Box_{\Mb[h]})G_{\Mb[h]} \omega_{\Euro^+}\right],
\]
where we have used the fact that $G^\textrm{ret}_{\Mb[h]} \omega_{\Euro^+}$ vanishes on the support of $h$. 
This expression is independent of the time-slice map $L^+$, because $\omega_{\Euro^+}$ is fixed modulo
the image of $\Box_\Mb$ on $2$-forms compactly supported in the image of $\imath^+_\Mb[h]$, on which 
$\Box_\Mb$ and $\Box_{\Mb[h]}$ agree. Standard manipulations with differential forms and the equivalence relation  give
\[
\rce^{\mathcal{F}\!_u}_{\Mb}\left[h\right]\left[\omega\right] =\left[\omega\right]
 -\left[\left(\delta_{\Mb\left[h\right]}-\delta_\Mb\right)G_{\Mb[h]}d_\Mb\omega_{\Euro^+}\right],
\]
for any $\left[\omega\right]\in\left[\Omega^2_0\left(M;\mathbb{K}\right)\right]$. Finally, the 
relative Cauchy evolution of $\textswab{F}_u$ is given by
\begin{align*}
\rce^{\textswab{F}_u}_\Mb\left[h\right]=\mathcal{Q}\left(\rce^{\mathcal{F}\!_u}_{\Mb}\left[h\right]\right).
\end{align*}

\subsection{The failure of dynamical locality for the universal theory}
In Subsection \ref{subsec relative Cauchy evolution universal F-theory}, we have already seen an example which shows that $\mathcal{F}\!_u$ and $\textswab{F}_u$ cannot possibly be dynamically local in the original sense of this definition~\cite{SPASs12}. 
To be more specific, let $\Nb\in\Loc$ be the Minkowski spacetime, $\Mb$ the Cauchy development in $\Nb$ of the set $\left\{0\right\}\times\left\{\left(x,y,z\right)^\top\in\mathbb{R}^3\mid x^2+y^2+z^2>1\right\}$ and $\psi:\Nb\to\Mb$ the inclusion map, 
then $\text{f}^{\,\kin}_{\,\Nb;M}=\mathcal{F}\!_u\psi:\mathcal{F}^{\kin}_u\left(\Nb;M\right)=\mathcal{F}\!_u\Mb\to\mathcal{F}\!_u\Nb$ fails to be 
monic, as we have seen, and therefore cannot be equivalent to the (necessarily monic)  
subobject $\text{f}^{\,\dyn}_{\,\Nb;M}:\mathcal{F}^{\dyn}_u\left(\Nb;M\right)\to\mathcal{F}\!_u\Nb$. 
Similarly, in the quantised case, the subobject $\varphi^{\dyn}_{\Nb;M}:\textswab{F}^{\dyn}_{\!u}\left(\Nb;M\right)\to\textswab{F}_u\Nb$ cannot be equivalent to the non-monic $\varphi^{\kin}_{\Nb;M}=\textswab{F}_u\psi:\textswab{F}_{\!u}^{\kin}\left(\Nb;M\right)=\textswab{F}_u\Mb\to\textswab{F}_u\Nb$. In this subsection, we show that the failure of dynamical locality for these theories is even more severe and cannot be achieved even if we
restrict to {\em contractible} globally hyperbolic open subsets. 

Let $\Mb\in\Loc$ be such that $H^2_{dR}\left(M;\mathbb{K}\right)\neq0$. 
By arguments given in Sect.~\ref{sec:univ fails}, there exists $\omega\in\Omega^2_0\left(M;\mathbb{K}\right)$ satisfying $d_\Mb\omega=0$ but $\left[\omega\right]\neq0\in\left[\Omega^2_0\left(M;\mathbb{K}\right)\right]$ (and hence $\widehat{\textbf{F}}_\Mb\left(\omega\right)\neq0\in\textswab{F}_u\Mb$). 
In other words, $[\omega]$ is a magnetic topological degeneracy.  
Lemma \ref{lem time-slice axiom 2} yields
\begin{flalign*}
&&\rce^{\mathcal{F}\!_u}_\Mb\left[h\right]\left[\omega\right]=\left[\omega\right]&&\makebox[0pt][r]{$\forall h\in H(\Mb)$,}
\end{flalign*}
and hence $\left(\rce^{\textswab{F}_u}_\Mb\left[h\right]\right)\left(\widehat{\textbf{F}}_\Mb\left(\omega\right)\right) =\widehat{\textbf{F}}_\Mb\left(\omega\right)$ for all $h\in H\left(\Mb\right)$.
Consequently, we have $\left[\omega\right]\in\mathcal{F}_{\!u}^\bullet\left(\Mb;K\right)$ and $\widehat{\textbf{F}}_\Mb\left(\omega\right)\in\textswab{F}_{\!u}^{\bullet}\left(\Mb;K\right)$ for all $K\in\mathscr{K}\left(\Mb;O\right)$ and for all contractible $O\in\mathscr{O}\left(\Mb\right)$. This implies $\left[\omega\right]\in\mathcal{F}_{\!u}^{\dyn}\left(\Mb;O\right)$ and $\widehat{\textbf{F}}_\Mb\left(\omega\right)\in\textswab{F}_{\!u}^{\dyn}\left(\Mb;O\right)$ for all contractible $O\in\mathscr{O}\left(\Mb\right)$. As $\left[\omega\right]$ is in the radical of the (complexified if $\mathbb{K}=\mathbb{C}$) pre-symplectic form on 
$\mathcal{F}\!_u\Mb$, it follows that $\mathcal{F}_{\!u}^{\dyn}\left(\Mb;O\right)$  has a degenerate (complexified if $\mathbb{K}=\mathbb{C}$) pre-symplectic form, while $\mathcal{F}_{\!u}^{\kin}\left(\Mb;O\right)$ is weakly non-degenerate; thus the subobject 
$\text{f}^{\,\dyn}_{\,\Mb;O}:\mathcal{F}_{\!u}^{\dyn}\left(\Mb;O\right)\to\mathcal{F}\!_u\Mb$ cannot possibly be equivalent to the subobject $\text{f}^{\,\kin}_{\,\Mb;O}:\mathcal{F}_{\!u}^{\kin}\left(\Mb;O\right)\to\mathcal{F}\!_u\Mb$ for any contractible $O\in\mathscr{O}\left(\Mb\right)$. The same is true for the quantised universal free F-theory because, for every contractible $O\in\mathscr{O}\left(\Mb\right)$, $\textswab{F}_{\!u}^{\dyn}\left(\Mb;O\right)$ is not simple, while $\textswab{F}_{\!u}^{\kin}\left(\Mb;O\right)$ is simple;
hence the subobjects $\varphi^{\dyn}_{\Mb;O}$ and $\varphi^{\kin}_{\Mb;O}$ are inequivalent. 
As far as the dynamical net is concerned, the elements $\left[\omega\right]$ resp. $\widehat{\textbf{F}}_\Mb\left(\omega\right)$, 
where $[\omega]$ is a magnetic topological degeneracy, are local to all regions. The same is true for the electric topological degeneracies, because electromagnetic
duality rotations through $\pi/2$ exchange electric and magnetic
topological
degeneracies and commute with relative Cauchy evolution, because they
are automorphisms of the theory~\cite[Prop.2.1]{Few13}. 
We summarise:

\begin{thm} The classical and the quantised universal free F-theory (and hence also the A-theory) are not dynamically local (even in the weakened sense obtained by restricting to contractible open globally hyperbolic subsets).
\end{thm}

\section{Dynamical locality of the reduced theory}\label{sec dynamical locality reduced theory}
In the last section we saw that the classical and the quantised universal free F-theory (and hence A-theory) fail local covariance and dynamical locality. However, we were also able to clearly identify what causes this failure, namely the possibility of having non-trivial radicals in the classical case and non-trivial centres in the quantum case. The reduced theories are free of these features and, as we will show, they are dynamically local. We work in the F-description, but all our statements have analogues in the equivalent A-description. 


\subsection{The relative Cauchy evolution of the reduced theory}
Having established local covariance, we will now show that the classical and the quantised reduced free 
F-theories obey the time-slice axiom. We will compute their respective relative Cauchy evolutions and differentiate them with respect to the metric perturbation, thus obtaining the stress-energy tensor for the classical reduced free F-theory. Since $\widetilde{\textswab{F}}=\mathcal{Q}\circ\widetilde{\mathcal{F}}$, we concentrate on the classical case.

The only difference to the subsections \ref{subsec time-slice axiom universal F-theory} and \ref{subsec relative Cauchy evolution universal F-theory} is so far the use of a different equivalence relation and hence different equivalence classes, i.e. $\L\cdot\R$ instead of $\left[\cdot\right]$. Assume $\psi\in\Loc\left(\Mb,\textbf{N}\right)$ is Cauchy and $L:\Omega^p_0\left(N;\mathbb{K}\right)\to\Omega^p_0\left(N;\mathbb{K}\right)$ is a time-slice map for $\left(\psi,\lambda^p_N,\Box_\textbf{N}\right)$. By Lem.~\ref{lem time-slice axiom 3},  $\omega_{\Euro}\in\iota_{\psi\left(M\right)*}\left(\Omega_{0,d}^p\left(\psi\left(M\right);\mathbb{K}\right)\oplus
\Omega_{0,\delta}^{p}\left(\psi\left(M\right);\mathbb{K}\right)\right)$ for $\omega\in\Omega^p_0\left(M;\mathbb{K}\right)$ such that $d_\Mb\omega=0$ or $\delta_\Mb\omega=0$. Thus we can adapt the results of Subsection \ref{subsec time-slice axiom universal F-theory} and Subsection \ref{subsec relative Cauchy evolution universal F-theory} by just replacing $\left[\cdot\right]$ with $\L\cdot\R$. In particular, $\widetilde{\mathcal{F}}$ and $\widetilde{\textswab{F}}$ obey the time-slice axiom and their respective relative Cauchy evolutions induced by $h\in H\left(\Mb\right)$ are given (in the same conventions as in Subsection \ref{subsec relative Cauchy evolution universal F-theory}; in particular,`$\Euro^+$' refers to an arbitrary time-slice map $L^+:\Omega^2_0\left(M;\mathbb{K}\right)\to\Omega^2_0\left(M;\mathbb{K}\right)$ for $\left(\imath^+_\Mb\left[h\right]:\Mb^+\left[h\right]\to\Mb,\lambda^2_M,\Box_\Mb\right)$)  by
\begin{flalign}\label{classical G-relative Cauchy evolution I}
&&\rce^{\widetilde{\mathcal{F}}}_{\Mb}\left[h\right]\omegaa=
\omegaa+\L\left(\delta_{\Mb\left[h\right]}-\delta_\Mb\right)G^\text{adv}_{\Mb\left[h\right]} d_\Mb\omega_{\Euro^+}\R
=\omegaa -\L\left(\delta_{\Mb\left[h\right]}-\delta_\Mb\right)G_{\Mb\left[h\right]} d_\Mb\omega_{\Euro^+}\R,&&\\\nonumber
&&&&\makebox[0pt][r]{$\omegaa\in\L\Omega^2_0\left(M;\mathbb{K}\right)\R$,} 
\end{flalign}
and also
\begin{flalign}\label{quantised G-relative Cauchy evolution}
&\rce^{\widetilde{\textswab{F}}}_\Mb\left[h\right]=\mathcal{Q}\left(\rce^{\widetilde{\mathcal{F}}}_{\Mb}\left[h\right]\right).
\end{flalign}
The intermediate expression in \eqref{classical G-relative Cauchy evolution I} allows us to employ a 
Born expansion as in \cite[(B.2)]{FeV12},
\begin{flalign*}
&&G_{\Mb\left[h\right]}^\text{adv}\omega=G^\text{adv}_\Mb\omega&-G^\text{adv}_\Mb\left(\Box_{\Mb\left[h\right]}-\Box_\Mb\right)G^\text{adv}_{\Mb\left[h\right]}\omega
&&&&&\makebox[0pt][r]{$\forall\omega\in\Omega^2_0\left(M;\mathbb{K}\right)$,}
\end{flalign*} 
in order to further compute:
\begin{flalign*}
\rce^{\widetilde{\mathcal{F}}}_{\Mb}\left[h\right]\omegaa&=\omegaa+\L\left(\delta_{\Mb\left[h\right]} -\delta_\Mb\right) G^\text{adv}_\Mb d_\Mb\omega_{\Euro^+}\R&\\
&\phantom{=\left[\omega\right]+}-\L\left(\delta_{\Mb\left[h\right]} -\delta_\Mb\right) G^\text{adv}_\Mb\left(\Box_{\Mb\left[h\right]}-\Box_\Mb\right) G^\text{adv}_{\Mb\left[h\right]}d_\Mb\omega_{\Euro^+}\R,&&&&&\makebox[0pt][r]{$\omegaa\in\L\Omega^2_0\left(M;\mathbb{K}\right)\R$}.
\end{flalign*} 
Now, $\supp G^\text{ret}_\Mb\omega_{\Euro^+}\cap\supp h=\emptyset$ by construction for any $\omega\in\Omega^2_0\left(M;\mathbb{K}\right)$ and, as $\delta_{\Mb\left[h\right]}-\delta_\Mb$ vanishes outside $\supp h$, we can thus replace $G^\text{adv}_\Mb d_\Mb\omega_{\Euro^+}$ by $-G_\Mb d_\Mb\omega_{\Euro^+}=-G_\Mb d_\Mb\omega$ to obtain
\begin{flalign}\label{classical G-relative Cauchy evolution II}
&&\rce^{\widetilde{\mathcal{F}}}_{\Mb}\left[h\right]\omegaa=\omegaa-\L\left(\delta_{\Mb\left[h\right]}-\delta_\Mb\right) \left(G_\Mb d_\Mb\omega  + G^\text{adv}_\Mb\left(\Box_{\Mb\left[h\right]}  -\Box_\Mb\right)G^\text{adv}_{\Mb\left[h\right]}d_\Mb\omega_{\Euro^+}\right)\R,&&\\\nonumber
&&&&\makebox[0pt][r]{$\omegaa\in\L\Omega^2_0\left(M;\mathbb{K}\right)\R$}.
\end{flalign}

For $\Mb\in\Loc$, we can associate to each $\omegaa\in\L\Omega^2_0\left(M;\mathbb{K}\right)\R$ a solution of the free Maxwell equations (\ref{eq. free Maxwell F}) with compact support on smooth spacelike Cauchy surfaces for $\Mb$ by setting $F_{\omegaa}:=d_\Mb G_\Mb\delta_\Mb\omega$ for any representative $\omega\in\Omega^2_0\left(M;\mathbb{K}\right)$. Clearly, all representatives will give rise to the same solution and if $d_\Mb G_\Mb\delta_\Mb\eta=F_{\omegaa}$ for $\etaa\in\L\Omega^2_0\left(M;\mathbb{K}\right)\R$, $\etaa=\omegaa$ necessarily. Thus, in the classical reduced free F-theory, we are only dealing with solutions of (\ref{initial value problem F}) which are of the form $d_\Mb G_\Mb\delta_\Mb\omega$ for $\omega\in\Omega^2_0\left(M;\mathbb{K}\right)$. Note that each solution of the Cauchy problem (\ref{initial value problem F}) will be of this form if $\Mb\in\Loc_\copyright$ (cf. Subsection \ref{subsec contractible classical phase space}). This provides a nice interpretation of the relative Cauchy evolution:
\begin{equation*}
d_{\Mb\left[h\right]}G_{\Mb\left[h\right]}\delta_{\Mb\left[h\right]}\left(\widetilde{\mathcal{F}}\jmath^+_\Mb\left[h\right]\right) \left(\left(\widetilde{\mathcal{F}}\imath^+_\Mb\left[h\right]\right)^{-1}\omegaa\right)=d_{\Mb\left[h\right]}G_{\Mb\left[h\right]} \delta_{\Mb\left[h\right]}\omega_{\Euro^+}
\end{equation*}
is the unique solution of the free Maxwell equations on $\Mb[h]$ which coincides with $F_{\omegaa}$ on $M^+[h]$ (cf. \cite{FeV12}). The agreement is not difficult to see, the uniqueness follows from the well-posedness of the Cauchy problem. Then, if $\eta\in\Omega^2_0\left(M;\mathbb{K}\right)$ is a representative of $\rce^{\widetilde{\mathcal{F}}}_{\Mb}\left[h\right]\omegaa$, then $d_\Mb G_\Mb\delta_\Mb\eta$ is the unique solution of the free Maxwell equations for the field strength on $\Mb$ agreeing with $d_{\Mb\left[h\right]}G_{\Mb\left[h\right]}\delta_{\Mb\left[h\right]} \omega_{\Euro^+}$ on $M^-\left[h\right]$. This interpretation of the relative Cauchy evolution will become very helpful in the proof of Lemma \ref{lem 1}.   

\subsection{The stress-energy tensor of the classical modifed theory}\label{subsec stress-energy tensor F-theory}
To show that $\widetilde{\mathcal{F}}$ and $\widetilde{\textswab{F}}$ are dynamically local, it will be helpful to
relate the relative Cauchy evolution to the stress-energy tensor for the classical reduced free F-theory. This can be done as follows: taking any compactly supported, symmetric and smooth tensor field\footnote{Recall, $\tau^*_M$ denotes the cotangent bundle of the smooth manifold $M$.} $h\in\Gamma^\infty_0\left(\tau_M^*\odot\tau_M^*\right)$, there is an interval $\left(-\varepsilon,\varepsilon\right)$ for some $\varepsilon>0$ such that $th\in H\left(\Mb\right)$ for all $t\in\left(-\varepsilon,\varepsilon\right)$ (cf.~\cite[\S\S 2\&3]{FeV12}). The relative Cauchy evolution for $\widetilde{\mathcal{F}}$ induced by $th\in H\left(\Mb\right)$ for $\Mb\in\Loc$ is differentiable in the weak symplectic topology (cf.~\cite[\S 3 \& Appx.B]{FeV12}), i.e.\ there is a $\mathbb{K}$-linear map $H_\Mb\left[h\right]:\widetilde{\mathcal{F}}\Mb\to\widetilde{\mathcal{F}}\Mb$ such that 
\begin{flalign}\label{stress-energy tensor 4}
&&\tilde{\textswab{w}}_\Mb\left(H_\Mb\left[h\right]\omegaa,\etaa\right)=\frac{d}{dt}\tilde{\textswab{w}}_\Mb\left(\rce^{\widetilde{\mathcal{F}}}_{\Mb}\left[th\right]\omegaa,\etaa\right)\Bigl|_{t=0}\,,&&\omegaa,\etaa\in\L\Omega^2_0\left(M;\mathbb{K}\right)\R
\end{flalign}
and the derivative on the right hand side exists for all such $\omegaa,\etaa$. Note, $H_\Mb\left[h\right]$ is called $F_\Mb\left[h\right]$ in \cite{FeV12}, a notation we avoid for obvious reasons. Inserting (\ref{classical G-relative Cauchy evolution II}) and already dropping some terms of order $t^2$ and higher, we need to compute (up to first order in $t$)
\begin{flalign}\label{stress-energy tensor 1}
&&\frac{d}{dt}\tilde{\textswab{w}}_\Mb\left(\rce^{\widetilde{\mathcal{F}}}_{\Mb}\left[th\right]\omegaa,\etaa\right)\Bigl|_{t=0}&=\lim_{t\rightarrow0}\tilde{\textswab{w}}_\Mb\left(\L-t^{-1}\left(\delta_{\Mb\left[th\right]}-\delta_\Mb\right)d_\Mb G_\Mb\omega\R,\etaa\right),&&\\\nonumber
&&&=-\lim_{t\rightarrow0}\int_Mt^{-1}\left(\delta_{\Mb\left[th\right]}-\delta_\Mb\right)d_\Mb G_\Mb\omega\wedge*d_\Mb G_\Mb\delta_\Mb\eta,&&\\\nonumber
&&&&&\makebox[0pt][r]{$\omegaa,\etaa\in\L\Omega^2_0\left(M;\mathbb{K}\right)\R$.}
\end{flalign}
The coderivative  $\delta_{\Mb\left[th\right]}$ may be expanded by a lengthy but straightforward 
computation (being careful to recall that the inverse metric to $g+th$ is  $\left(g+th\right)^{-1}=g^{-1}-th^{\sharp\sharp}+O(t^2)$, which reads in abstract index notation $g^{ab}-t h^{ab}+O(t^2)$):
\begin{flalign*}
&&\left(\left(\delta_{\Mb\left[th\right]}-\delta_\Mb\right)\varpi\right)_{cd}= t\left(\nabla\!_a\left(h^{ab}\varpi_{bcd}\right)-\frac{1}{2}\left(\nabla_bh^a_a\right)\varpi^b_{\phantom{b}cd}+\left(\nabla\!_ah_{bc}\right)\varpi^{ab}_{\phantom{ab}d} -\left(\nabla\!_ah_{bd}\right)\varpi^{ab}_{\phantom{ab}c}\right)+O(t^2),&&\\
&&&&\makebox[0pt][r]{$\varpi\in\Omega^3\left(M;\mathbb{K}\right)$,}
\end{flalign*}   
where $\nabla$ stands for the Levi-Civita connection with respect to $g$. This yields
\begin{flalign}\label{stress-energy tensor 2}
&&H_\Mb\left[h\right]\L\omega_{cd}\R&=\!\L-\nabla\!_a\left(h^{ab}\left(G_\Mb d_\Mb\omega\right)_{bcd}\right)+\frac{1}{2}\left(\nabla_bh^a_a\right)\left(G_\Mb d_\Mb\omega\right)^b_{\phantom{b}cd}\right.&&\\\nonumber
&&&\phantom{=}-\left.\left(\nabla\!_ah_{bc}\right)\left(G_\Mb d_\Mb\omega\right)^{ab}_{\phantom{ab}d}+\left(\nabla\!_ah_{bd}\right)\left(G_\Mb d_\Mb\omega\right)^{ab}_{\phantom{ab}c}\R,&&\omegaa\in\L\Omega^2_0\left(M;\mathbb{K}\right)\R,
\end{flalign}
whose well-definedness can be seen by using the weak non-degeneracy of $\tilde{\textswab{w}}_\Mb$. In order to bring (\ref{stress-energy tensor 1}) into a nicer form, we define $\varpi:=d_\Mb G_\Mb\omega\in\Omega^3(M;\mathbb{K})$ and $F_{\etaa}:=d_\Mb G_\Mb\delta_\Mb\eta$. The divergence theorem entails the following identities
\begin{align*}
\int_M\nabla\!_a\left(h^{ab}\varpi_{bcd}\right)F_{\etaa}^{cd}\vol_\Mb=-\int_Mh^{ab}\varpi_{bcd}\nabla\!_aF_{\etaa}^{cd}\vol_\Mb,
\end{align*}
\begin{align*}
\int_M\left(\nabla\!_bh^a_a\right)\varpi^b_{\phantom{b}cd}F_{\etaa}^{cd}\vol_\Mb&=-\int_Mh^a_a\nabla\!_b\left(\varpi^b_{\phantom{b}cd}\right)F_{\etaa}^{cd}\vol_\Mb-\int_Mh^a_a\varpi^b_{\phantom{b}cd}\nabla\!_bF_{\etaa}^{cd}\vol_\Mb
\end{align*}
and
\begin{align*}
\int_M\left(\nabla\!_ah_{bc}\right)\varpi^{ab}_{\phantom{ab}d}F_{\etaa}^{cd}\vol_\Mb=-\int_Mh^b_c\left(\nabla\!_a\varpi^a_{\phantom{a}bd}\right)F_{\etaa}^{cd})\vol_\Mb-\int_Mh_{bc}\varpi^{ab}_{\phantom{ab}d}\nabla\!_aF_{\etaa}^{cd}\vol_\Mb,
\end{align*}
where $\nabla\!_b\left(\varpi^b_{\phantom{b}cd}\right)=-\left(\delta_\Mb\varpi\right)_{cd}=+\left(d_\Mb G_\Mb\delta_\Mb\omega\right)_{cd}=: F_{\omegaa\!cd}$; together with
$d_\Mb F_{\etaa} = 0$ and 
\begin{align*}
\varpi^b_{\phantom{b}cd}\nabla\!_bF_{\etaa}^{cd}\vol_\Mb=\varpi_{bcd}\nabla\!^{[b}F_{\etaa}^{cd]}\vol_\Mb=
3!\varpi\wedge * d_\Mb F_{\etaa} = 0,
\end{align*}
they yield overall
\begin{flalign*}
&&\tilde{\textswab{w}}_\Mb\left(H_\Mb\left[h\right]\omegaa,\etaa\right)&=\frac{d}{dt}\tilde{\textswab{w}}_\Mb\left(\rce^{\widetilde{\mathcal{F}}}_{\Mb}\left[h\right]\omegaa,\etaa\right)\Bigl|_{t=0}&&\\
&&&=-\int_Mh_{ab}\left(\frac{1}{4}g^{ab}F_{\omegaa\!mn}F_{\etaa}^{mn}-g_{mn}F_{\omegaa}^{am}F_{\etaa}^{bn}\right)\vol_\Mb&&\\
&&&=-\int_Mh_{ab} T^{ab}_\Mb\left(\omegaa,\etaa\right)\vol_\Mb,&&\makebox[0pt][r]{$\omegaa,\etaa\in\L\Omega^2_0\left(M;\mathbb{K}\right)\R$.}
\end{flalign*}
(Note that there is a sign error in the analogous formula \cite[Eq.~(3.7)]{FeV12}, which however does not alter the main
results of that reference.)
Here $T_\Mb\left(\omegaa,\etaa\right)$ is the polarised form of the stress-energy tensor for the classical reduced free F-theory on $\Mb\in\Loc$ 
\begin{flalign}\label{stress-energy tensor 3}
&&T^{ab}_\Mb\left(\omegaa,\etaa\right)=\frac{1}{4}g^{ab}F_{\omegaa\!mn}F_{\etaa}^{mn}-g_{mn}F_{\omegaa}^{am}F_{\etaa}^{bn},&&\omegaa,\etaa\in\L\Omega^2_0\left(M;\mathbb{K}\right)\R,
\end{flalign}
where $F_{\omegaa}:=d_\Mb G_\Mb\delta_\Mb\omega$ with a representative $\omega\in\Omega^2_0\left(M;\mathbb{K}\right)$ for $\omegaa\in\L\Omega^2_0\left(M;\mathbb{K}\right)\R$ (one could also regard this expression
as half of a second directional derivative of the stress-energy tensor). Note, the same expression (\ref{stress-energy tensor 3}) is obtained for the stress-energy tensor of the classical universal free F-theory if $\L\cdot\R$ is replaced with $\left[\cdot\right]$.

\subsection{Verification of dynamical locality for the reduced theories}
We will now prove that the reduced free F-theory $\widetilde{\mathcal{F}}:\Loc\to\SymplK$ obeys dynamical locality
(hence the same is true for the corresponding reduced A-theory). In order for equalisers, unions and intersections to exist, we regard $\widetilde{\mathcal{F}}$ as a functor $\widetilde{\mathcal{F}}:\Loc\to\mpSymplK$. We will follow the reasoning of \cite{FeV12} using the stress-energy tensor of $\widetilde{\mathcal{F}}$ in order to characterise the dynamical net. The main technical point of difference is that the field strength tensor satisfies not only the wave equation but also the free Maxwell equations. 

\begin{lem}\label{lem 1} Let $K$ be any compact subset of $\Mb\in\Loc$. Then 
\begin{align}\label{eq:lem1}
\widetilde{\mathcal{F}}^\bullet\left(\Mb;K\right)=\left\{\omegaa\in\widetilde{\mathcal{F}}\Mb\mid\supp T_\Mb\left(\omegaa,\overline{\omegaa}\right)\subseteq J_\Mb\left(K\right)\right\}=\bigcap_{\substack{h\in\Gamma^\infty_0\left(\tau_\Mb^*\odot\tau_\Mb^*\right)\\\supp h\subseteq K^\perp}}\ker H_\Mb\left[h\right],
\end{align}
and also $\widetilde{\mathcal{F}}^\bullet\left(\Mb;K\right)=
\left\{\omegaa\in\widetilde{\mathcal{F}}\Mb\mid\supp F_{\omegaa} \subseteq J_\Mb\left(K\right)\right\}$.
\end{lem}
\noindent\textbf{\textit{Proof:}} Labelling the members of \eqref{eq:lem1} as 
I, II and III respectively, we will prove that 
$\textrm{I}\subseteq \textrm{III}\subseteq \textrm{II}\subseteq\textrm{I}$. 
Starting with $\textrm{I}\subseteq \textrm{III}$, suppose $\omegaa\in\widetilde{\mathcal{F}}^\bullet\left(\Mb;K\right)$. For $h\in\Gamma^\infty_0\left(\tau_\Mb^*\odot\tau_\Mb^*\right)$ with support $\supp h\subseteq K^\perp$, there is $\varepsilon>0$ such that $th\in H\left(\Mb;K^\perp\right)$ for all $t\in\left(-\varepsilon,\varepsilon\right)$. As $\rce^{\widetilde{\mathcal{F}}}_\Mb\left[th\right]\omegaa=\omegaa$ for all $t\in\left(-\varepsilon,\varepsilon\right)$, we have $\frac{d}{dt}\tilde{\textswab{w}}_\Mb\left(\rce^{\widetilde{\mathcal{F}}}_\Mb\left[th\right]\omegaa,\etaa\right)\bigl|_{t=0}=0$ for all $\etaa\in\widetilde{\mathcal{F}}\Mb$. Hence also $\tilde{\textswab{w}}_\Mb\left(H_\Mb\left[h\right]\omegaa,\etaa\right)=0$
for all $\etaa\in\widetilde{\mathcal{F}}\Mb$ and so by weak non-degeneracy, 
$\omegaa\in\ker H_\Mb\left[h\right]$; as $h$ was arbitrary, we have $\textrm{I}\subseteq \textrm{III}$. 
For $\textrm{III}\subseteq \textrm{II}$, if
\[
\omegaa\in\bigcap_{\substack{h\in\Gamma^\infty_0\left(\tau_\Mb^*\odot\tau_\Mb^*\right)\\\supp h\subseteq K^\perp}}\ker H_\Mb\left[h\right],
\]
then, in particular, 
 $\tilde{\textswab{w}}_\Mb\left(H_\Mb\left[h\right]\omegaa,\overline{\omegaa}\right)=-\int_Mh_{ab}T^{ab}_\Mb\left(\omegaa,\overline{\omegaa}\right)\vol_\Mb=0$ for all $h\in\Gamma^\infty_0\left(\tau_\Mb^*\odot\tau_\Mb^*\right)$ with support $\supp h\subseteq K^\perp$,
so $\supp T_\Mb\left(\omegaa,\overline{\omegaa}\right)\subseteq J_\Mb\left(K\right)$ as required. 
Finally, to prove $\textrm{II}\subseteq\textrm{I}$, we note that  $\supp T_\Mb\left(\omegaa,\overline{\omegaa}\right)\subseteq J_\Mb\left(K\right)$ implies that $\supp \left(F_{\omegaa}\right)\subseteq J_\Mb\left(K\right)$ because the energy density, which is the sum of the squares of the off-diagonal components of $F_{\omegaa}$ (in some local framing), must vanish at each point $p\not\in J_\Mb\left(K\right)$. Accordingly, $F_{\omegaa}$ is a solution of Maxwell's equations in the perturbed spacetime $\Mb\left[h\right]$ for every $h\in H\left(\Mb;K^\perp\right)$. Hence, it is the unique solution on $\Mb\left[h\right]$ that coincides with $F_{\omegaa}$ on $M^+\left[h\right]$ and also the unique solution on $\Mb$ that coincides with $F_{\omegaa}$ on $M^-\left[h\right]$. Thus, $\omegaa$ and $\rce^{\widetilde{\mathcal{F}}}_{\Mb}\left[h\right]\omegaa$ give rise to the same solution of the free Maxwell equations on $\Mb$ which implies $\rce^{\widetilde{\mathcal{F}}}_{\Mb}\left[h\right]\omegaa=\omegaa$ 
and consequently $\omegaa\in\widetilde{\mathcal{F}}^\bullet\left(\Mb;K\right)$.
The final statement is immediate from the argument just given.
\hfill\SquareCastShadowTopRight


\begin{lem}\label{lem 2} For all $O\in\mathscr{O}\left(\Mb\right)$, there is a subobject $m_O:\widetilde{\mathcal{F}}^{\kin}\left(\Mb;O\right)\to\widetilde{\mathcal{F}}^{\dyn}\left(\Mb;O\right)$ such that $\tilde{\emph{f}}^{\,\dyn}_{\Mb;O}\circ m_O=\tilde{\emph{f}}^{\,\kin}_{\Mb;O}$ holds for the subobjects $\tilde{\emph{f}}^{\,\dyn}_{\Mb;O}:\widetilde{\mathcal{F}}^{\dyn}\left(\Mb;O\right)\to\widetilde{\mathcal{F}}\Mb$ and $\tilde{\emph{f}}^{\,\kin}_{\Mb;O}:\widetilde{\mathcal{F}}^{\kin}\left(\Mb;O\right)\to\widetilde{\mathcal{F}}\Mb$. 
\end{lem}
\noindent\textbf{\textit{Proof:}} Let $\omegaa\in\widetilde{\mathcal{F}}^{\kin}\left(\Mb;O\right)=\widetilde{\mathcal{F}}\left(\Mb|_O\right)$ and $\omega\in\Omega^2_0\left(O;\mathbb{K}\right)$ a representative of $\omegaa$. 
Choosing for each $x\in\supp\omega$ a Cauchy ball $B_x$ containing $x$ and taking the Cauchy developments, we have found an open cover $\left\{D_\Mb(B_x)\right\}_{x\in\supp\omega}$ of $\supp\omega$ in $\Mb$. Since $\supp\omega$ is compact, finitely many of these sets are enough to cover $\supp\omega$, say $\supp\omega\subseteq\bigcup_{i=0}^nD_\Mb\left(B_i\right)$ with $n\geq0$. 

Let $\left\{\chi,\chi^i\mid i=0,\dots,n\right\}$ be a smooth partition of unity subordinated to the open cover $\left\{M\setminus\supp\omega, D_\Mb\left(B_i\right)\mid i=0,\dots,n\right\}$ of $M$. Defining for all $i\in I$ $\omega_i:=\chi^i\iota_{O*}\omega\in\Omega^2_0\left(M;\mathbb{K}\right)$ with $\supp\omega_i\subseteq D_\Mb\left(B_i\right)\cap O$, we can write $\iota_{O*}\omega=\sum_{i=0}^n\omega_i$. By construction, $\supp\omega_i\in\mathscr{K}\left(\Mb;O\right)$. As $\supp T_\Mb\left(\L\omega_i\R,\overline{\L \omega_i\R}\right)\subseteq J_\Mb\left(\supp\omega_i\right)$, Lemma \ref{lem 1} yields $\L\omega_i\R\in\widetilde{\mathcal{F}}^\bullet\left(\Mb;\supp\omega_i\right)$ and hence, $\L\iota_{O*}\omega\R=\sum_{i=0}^n\L\omega_i\R\in\widetilde{\mathcal{F}}^{\dyn}\left(\Mb;O\right)$ because $\widetilde{\mathcal{F}}^{\dyn}\left(\Mb;O\right)$ is the smallest (compexified if $\mathbb{K}=\mathbb{C}$) pre-symplectic subspace of $\widetilde{\mathcal{F}}\Mb$ containing $\widetilde{\mathcal{F}}^\bullet\left(\Mb;K\right)$ for all $K\in\mathscr{K}\left(\Mb;O\right)$. Evidently, $m_O:\widetilde{\mathcal{F}}^{\kin}\left(\Mb;O\right)\to\widetilde{\mathcal{F}}^{\dyn}\left(\Mb;O\right)$ defined by $m_O\omegaa:=\L\iota_{O*}\omega\R$ for $\omegaa\in\widetilde{\mathcal{F}}^{\kin}\left(\Mb;O\right)$ is the subobject with the claimed property.\hfill\SquareCastShadowTopRight\par\bigskip

The following lemma can be considered as an analogue to \cite[Lem.3.1.]{FeV12} and is integral to the proof that the kinematic and the dynamic nets coincide.

\begin{lem}\label{lem 3} Let $\Mb\in\Loc$ and $K\subseteq O\in\mathscr{O}\left(\Mb\right)$ compact. There exists $\chi\in\mathcal{C}^\infty(M)$ such that every
solution $F\in\Omega^2\left(M,\mathbb{K}\right)$ of Maxwell's equations with $\supp F\subseteq J_\Mb\left(K\right)$ can be written as $F=G_\Mb\Box_\Mb\chi F$,
where $\Box_\Mb\chi F\in\Omega^2_0\left(M;\mathbb{K}\right)$, $\delta_\Mb\chi F\in\Omega^1_0\left(M,\mathbb{K}\right)$ and $d_\Mb\chi F\in\Omega^3_0\left(M,\mathbb{K}\right)$ are supported in $O$.  
\end{lem}
\noindent\textbf{\textit{Proof:}} The proof works in exactly the same way as that of \cite[Lem.3.1(i)]{FeV12}. The additional point is that due to $d_\Mb F=0$ and $\delta_\Mb F=0$, the Leibniz rule gives $d_\Mb\chi F=0$ and $\delta_\Mb\chi F=0$ outside of the compact set $K_0\subseteq O$ defined in \cite[Lem.3.1(i)]{FeV12}, and are thereby compactly supported in $O$.\hfill\SquareCastShadowTopRight\par\bigskip

Recall from Subsection \ref{subsec dynamical net} that for $\Mb\in\Loc$ and $O\in\mathscr{O}\left(\Mb\right)$, $\widetilde{\mathcal{F}}^{\dyn}_{\Mb;O}$ is the (complexified if $\mathbb{K}=\mathbb{C}$) pre-symplectic subspace of $\widetilde{\mathcal{F}}\Mb$ generated by $\bigcup_{K\in\mathscr{K}\left(\Mb;O\right)}\widetilde{\mathcal{F}}^{\bullet}\left(\Mb;K\right)$.

\begin{lem}\label{lem 4} For all $O\in\mathscr{O}\left(\Mb\right)$, there is a subobject $\mu_O:\widetilde{\mathcal{F}}^{\dyn}\left(\Mb;O\right)\to\widetilde{\mathcal{F}}^{\kin}\left(\Mb;O\right)$ such that $\tilde{\emph{f}}^{\,\kin}_{\Mb;O}\circ\mu_O=\tilde{\emph{f}}^{\,\dyn}_{\Mb;O}$ holds for the subobjects $\tilde{\emph{f}}^{\,\kin}_{\Mb;O}:\widetilde{\mathcal{F}}^{\kin}\left(\Mb;O\right)\to\widetilde{\mathcal{F}}\Mb$ and $\tilde{\emph{f}}^{\,\dyn}_{\Mb;O}:\widetilde{\mathcal{F}}^{\dyn}\left(\Mb;O\right)\to\widetilde{\mathcal{F}}\Mb$.
\end{lem}
\noindent\textbf{\textit{Proof:}} We start by showing that for each $K\in\mathscr{K}\left(\Mb;O\right)$, $\omegaa\in\widetilde{\mathcal{F}}^\bullet\left(\Mb;K\right)$ has a representative $\eta\in\Omega^2_0\left(M;\mathbb{K}\right)$ with $\supp\eta\subseteq O$. By Lemma \ref{lem 1}, we have $\supp d_\Mb G_\Mb\delta_\Mb\omega\subseteq J_\Mb\left(K\right)$ for any representative $\omega\in\Omega^2_0\left(M;\mathbb{K}\right)$ of $\omegaa$. 
Now, by definition of $\mathscr{K}\left(\Mb;O\right)$, 
$K$ has a neighbourhood comprising finitely many causally disjoint diamonds $\left\{D_\Mb\left(B_i\right)\right\}_{i=0,\dots,n}$, $n\geq0$, based in smooth spacelike Cauchy surfaces for $\Mb$ such that the bases $\left\{B_i\right\}_{i=0,\dots,n}$ are contained in $O$. Note that these diamonds might not be entirely contained in $O$. Hence, $\left\{D_{\Mb|_O}\left(B_i\right)\right\}_{i=0,\dots,n}$ are globally hyperbolic open subsets of both $\Mb|_O$ and $\Mb$, which are furthermore contractible. Because of the causal disjointness, their (disjoint) union $U:=\bigsqcup_{i=0}^nD_{\Mb|_O}\left(B_i\right)$ is a globally hyperbolic open subset of $\Mb|_O$ and $\Mb$, contains\footnote{$D_{\Mb|_O}\left(B_i\right)=D_\Mb\left(B_i\right)\cap O$ for $i=0,\dots,n$ because $O$ is causally convex in $\Mb$.} $K$ and each connected component is contractible. We apply Lemma \ref{lem 3} to $U$ and find that $F:=d_\Mb G_\Mb\delta_\Mb\omega=G_\Mb\Box_\Mb\chi F=-G_\Mb\delta_\Mb d_\Mb\chi F-G_\Mb d_\Mb\delta_\Mb\chi F$, where $d_\Mb\chi F\in\Omega^3_0\left(M;\mathbb{K}\right)$ and $\delta_\Mb\chi F\in\Omega^1_0\left(M;\mathbb{K}\right)$ are compactly supported in $U$. Since each connected component of $U$ is contractible, there are $\eta_1,\eta_2\in\Omega^2_0\left(U;\mathbb{K}\right)$ satisfying the equalities $d_\Mb\chi F=d_\Mb\iota_{U*}\eta_1$ and $\delta_\Mb\chi F=\delta_\Mb\iota_{U*}\eta_2$. Thus, $d_\Mb G_\Mb\delta_\Mb\omega=d_\Mb G_\Mb\delta_\Mb\iota_{U*}\left(\eta_1-\eta_2\right)$, which shows $\omegaa=\L\iota_{U*}\left(\eta_1-\eta_2\right)\R$. Accordingly,  $\eta:=\iota_{U*}\left(\eta_1-\eta_2\right)\in\Omega^2_0\left(M;\mathbb{K}\right)$ is a representative of $\omegaa$ that is compactly supported in $O$ (because $\eta$ is compactly supported in $U\subseteq O$).

A subobject $\mu_O:\widetilde{\mathcal{F}}^{\dyn}\left(\Mb;O\right)\to\widetilde{\mathcal{F}}^{\kin}\left(\Mb;O\right)$ is now defined by $\mu_O\omegaa:=\L\iota^*_O\eta\R$ for $\omegaa\in\widetilde{\mathcal{F}}^{\dyn}\left(\Mb;O\right)$, where $\eta\in\Omega^2_0\left(M;\mathbb{K}\right)$ is any representative of $\omegaa$ supported in $O$. $\mu_O$ is well-defined because if $\eta,\eta'\in\Omega^2_0\left(M;\mathbb{K}\right)$ are two representatives of $\omegaa\in\widetilde{\mathcal{F}}^{\dyn}\left(\Mb;\mathbb{K}\right)$ with compact support in $O$, $d_{\Mb|_O}G_{\Mb|_O}\delta_{\Mb|_O}\iota^*_O\left(\eta-\eta'\right)=d_{\Mb|_O}\iota^*_OG_\Mb\iota_{O*}\delta_{\Mb|_O}\iota^*_O\left(\eta-\eta'\right)=0$ and so $\iota^*_O\left(\eta-\eta'\right)=\alpha+\beta$ with $\alpha\in\Omega^2_{0,d}\left(O;\mathbb{K}\right)$ and $\beta\in\Omega^2_{0,\delta}\left(O,\mathbb{K}\right)$ which is equivalent to say $\L\iota^*_O\eta\R=\L\iota^*_O\eta'\R$. Clearly, $\tilde{\text{f}}^{\,\kin}_{\Mb;O}\circ\mu_O=\tilde{\text{f}}^{\,\dyn}_{\Mb;O}$, which shows that $\mu_O$ is a monic.\hfill\SquareCastShadowTopRight\par\bigskip

Combining Lemma \ref{lem 2} and Lemma \ref{lem 4}, the main statement of this subsection follows:
\begin{thm}\label{thm 1} The classical reduced theory of the free Maxwell field is dynamically local.
\end{thm}
\noindent\textbf{\textit{Proof:}} Let $\Mb\in\Loc$ and $O\in\mathscr{O}\left(\Mb\right)$ be arbitrary. Combining Lemmas \ref{lem 2} and \ref{lem 4}, it follows that
$\tilde{\text{f}}^{\,\kin}_{\Mb;O}$ and $\tilde{\text{f}}^{\,\dyn}_{\Mb;O}$ are equivalent subobjects and the classical reduced free F-theory $\widetilde{\mathcal{F}}:\Loc\to\mpSymplK$ is dynamically local. Due to natural isomorphism, this is also the case for the classical reduced free A-theory.\hfill\SquareCastShadowTopRight\par\bigskip

From Theorem \ref{thm 1}, we may deduce that the quantised reduced free F-theory $\widetilde{\textswab{F}}:\Loc\to\umAlg$ (and hence the quantised reduced free A-theory) is dynamically local:
\begin{cor} The quantised reduced theory of the free Maxwell field obeys dynamical locality.
\end{cor}
\noindent\textbf{\textit{Proof:}} $\widetilde{\textswab{F}}=\mathcal{Q}\circ\widetilde{\mathcal{F}}$ with the quantisation functor $\mathcal{Q}:\mpSymplK\to\umAlg$ and as a result of that we need to check $(\mathscr{L}1-\mathscr{L}4)$ of \cite[p.1688]{FeV12}:
\begin{enumerate}
\item[($\mathscr{L}1$)] The relative Cauchy evolution of $\widetilde{\mathcal{F}}$ is differentiable in the weak symplectic topology as in (\ref{stress-energy tensor 4}), and the resulting maps obey 
(the sign appears incorrectly in~\cite{FeV12})
\begin{flalign*}
&&\tilde{\textswab{w}}_\Mb\left(H_\Mb\left[h\right]\omegaa,\overline{\omegaa}\right)=-\int_Mh_{ab} T^{ab}_\Mb\left(\omegaa,\overline{\omegaa}\right)\vol_\Mb,&&\\
&&&&\makebox[0pt][r]{$\omegaa\in\L\Omega^2_0\left(M;\mathbb{K}\right)\R$, $h\in H\left(\Mb;O\right)$, $O\in\mathscr{O}\left(\Mb\right)$, $\Mb\in\Loc$,}
\end{flalign*}
where $T_\Mb\left(\omegaa,\overline{\omegaa}\right)\in\Gamma^\infty\left(\tau_\Mb\odot\tau_\Mb\right)$ for each $\omegaa\in\L\Omega^2_0\left(M;\mathbb{K}\right)\R$ and $\Mb\in\Loc$.

\item[($\mathscr{L}2$)] For each $O\in\mathscr{O}\left(\Mb\right)$ containing $\supp h$ of $h\in\Gamma^\infty_0\left(\tau^*_\Mb\odot\tau^*_\Mb\right)$, $\img H_\Mb\left[h\right]$ can be identified with a subset of $\widetilde{\mathcal{F}}^{\kin}\left(\Mb;O\right)$.

\item[($\mathscr{L}3$)] $\widetilde{\mathcal{F}}$ obeys extended locality, i.e. $\img\tilde{\text{f}}^{\,\kin}_{\Mb;O_1}\cap\img\tilde{\text{f}}^{\,\kin}_{\Mb;O_2}=0\in\widetilde{\mathcal{F}}\Mb$ for spacelike separated $O_1,O_2\in\mathscr{O}\left(\Mb\right)$, $\Mb\in\Loc$. 

\item[($\mathscr{L}4$)] $\widetilde{\mathcal{F}}^\bullet\left(\Mb;K\right)=\bigcap_{\substack{h\in\Gamma^\infty_0\left(\tau_\Mb^*\odot\tau_\Mb^*\right)\\\supp h\subseteq K^\perp}}\ker H_\Mb\left[h\right]$ for $K$ compact in $\Mb\in\Loc$.
\end{enumerate}
($\mathscr{L}1$) is obvious from what was done in Subsection \ref{subsec stress-energy tensor F-theory}. For $\Mb\in\Loc$, the image of $H_\Mb\left[h\right]$, where $h\in\Gamma^\infty_0\left(\tau^*_\Mb\odot\tau^*_\Mb\right)$, can be identified with a subset of $\widetilde{\mathcal{F}}^{\kin}\left(\Mb;O\right)$ for each $O\in\mathscr{O}\left(\Mb\right)$ with $\supp h\subseteq O\in\mathscr{O}\left(\Mb\right)$ by (\ref{stress-energy tensor 2}). ($\mathscr{L}3$) is obvious and $(\mathscr{L}4)$ is proven by Lemma \ref{lem 1}. Hence, \cite[Thm.5.3]{FeV12}\footnote{The sign error in \cite{FeV12} does not affect the validity of this result because the focus is on solutions with vanishing stress-energy tensor.} applies and proves the result.
\hfill\SquareCastShadowTopRight

\section{Discussion}\label{sec conclusions}

\subsection{Summary}

In this paper, we have discussed the notion of dynamical locality for the free Maxwell field. Describing the quantum field theory in terms of the universal algebra of the unital $*$-algebras of smeared quantum fields (cf. \cite{DL12}), and describing the classical field theory by the equivalent for  (complexified if $\mathbb{K}=\mathbb{C}$)  pre-symplectic spaces, we showed that the classical and the quantised universal theories, given by functors $\mathcal{F}\!_u,\mathcal{A}_u:\Loc\to\pSymplK$ and $\textswab{F}_u,\textswab{A}_u:\Loc\to\umAlg$, fail dynamical locality due to $\Loc$-objects $\Mb$ with $H^2_{dR}\left(M;\mathbb{K}\right)\neq0$. However, we were able to modify the classical and the quantised universal $F$-theory to obtain locally covariant and dynamically local theories $\widetilde{\mathcal{F}}:\Loc\to\SymplK$ and $\widetilde{\textswab{F}}:\Loc\to\umAlg$. In establishing this, we have used the same chain of arguments as \cite{FeV12} for the free real scalar field.

We have also found a generalisation of the ``natural algebraic relation'' in \cite{Bong77}, by means of  natural isomorphisms between $\mathcal{F}\!_u$ and $\mathcal{A}_u$, and between $\textswab{F}_u$ and $\textswab{A}_u$.  Hence, none of the theories $\mathcal{F}\!_u$, $\mathcal{A}_u$, $\textswab{F}_u$ and $\textswab{A}_u$ can accommodate observables relevant to the Aharonov-Bohm effect that are parameterised by the third compact support de Rham cohomology \cite{DHS12}. On the
other hand, all the theories discussed admit electromagnetic duality rotations as global symmetries.
To conclude, we discuss three aspects in more detail, namely the status of dynamical locality, the 
categorical structure underlying some of our constructions, and the relation of our present
work to the discussions of SPASs in~\cite{SPASs12,FeV12}. 

\subsection{Dynamical locality}

It is useful to summarise the current state of knowledge regarding dynamical
locality. For the Klein--Gordon theory in spacetime dimension $n\ge 2$, with mass $m$ and curvature coupling $\xi$, the theory is known to be dynamically local provided at least one of $m$ or $\xi$ is non-zero \cite{FeV12,Fer13}. The same is known to be true for the extended theory of Wick polynomials  for $m>0$ in the two cases of minimal and conformal coupling in 
dimensions $n\ge 2$ \cite{Fer13};  moreover, the Dirac field in $n=4$ dimensions
is dynamically local for $m\ge 0$ \cite{Fer13b}.

The massless minimally coupled scalar field fails to be dynamically
local in all dimensions $n\ge 2$, which can be traced to the rigid gauge symmetry $\phi\mapsto\phi+\text{const}$ of the theory; as mentioned, dynamical locality is restored if either $m$ or $\xi$ become non-zero. Moreover, the free massless current is also dynamically local
in dimensions $n\ge 3$, and also in $n=2$ if we restrict to the category of connected spacetimes~\cite{FeV12}. The inhomogeneous  minimally coupled Klein--Gordon theory has recently been studied \cite{FeSch14}; here, the category of spacetimes $\Loc$ is replaced by a
category of spacetimes with sources, and one modifies the definition of the relative Cauchy evolution and the dynamical net to take account of both metric and source perturbations. 
The result is that the inhomogeneous theory is dynamically local for all $n\ge 2$ and $m\ge 0$. 
Thus we see that the failure of dynamical locality is lifted as soon interactions, in the 
form of curvature coupling or external sources (or, mass terms) are included. Note that,
while the curvature and mass terms break the gauge symmetry, this is not the case
for the inhomogeneous theory.\footnote{There is a subtlety in \cite{FeSch14}: not all generators of relative Cauchy evolutions correspond to observable (gauge-invariant) fields in the $m=0$ case; if one excludes such relative Cauchy evolutions from the construction of
the dynamical net, then dynamical locality fails.} 

Our present results on the Maxwell field contribute to the emerging picture as follows. The failure of dynamical locality for the universal theory can
be traced to the existence of topological charges present whenever
the second de Rham cohomology is non-trivial. These observables are
invariant under all relative Cauchy evolutions and so are common to 
every element of the dynamical net, which does not distinguish
between observables that are local to every region and `observables that
are localised at infinity'. Actually, these observables can have
unusual spatial localisation as well: it is possible for such an element 
to be common to spacelike separated elements of the kinematic net, 
giving a failure of extended locality~\cite{Schoch1968,Landau1969}. 
In the quantum field theory, the topological charges are central elements
which parameterise different superselection sectors of the theory~\cite{AS80}, again underlining their global nature. By contrast, 
the reduced Maxwell theory in $n=4$ dimensions provides a well-behaved locally covariant and dynamically local theory (at the cost of
giving up topological observables labelled by de Rham cohomology
$H^p_{0,dR}$ for $p=1,2$). Overall, dynamical locality appears
to be a reasonable expectation for theories of local observables, 
but to fail where theories admit observables of an essentially global
nature that are stabilised by topological or other constraints. 

\subsection{Categorical structures}
A number of ideas concerning the `universal' and the `reduced' theory for the classical and the quantised free Maxwell field can be put in a broader categorical context. The details of the following discussion have been worked out and will appear in B.L.'s forthcoming Ph.D. thesis.

For each $\Mb=\left(M,g,\ogth,\tgth\right)\in\Loc$, we can consider the category $\mathcal{J}_\textbf{M}$ whose objects are those $\textbf{L}=\left(L,g_L,\ogth_L,\tgth_L\right)\in\Loc_\copyright$ such that $L\subseteq M$ is open and causally convex (excluding $L=M$ if $\textbf{M}\in\Loc_\copyright$), $g_L=g|_L$, $\ogth_L=\ogth|_L$ and $\tgth_L=\tgth|_L$; the morphisms in $\mathcal{J}_\textbf{M}$ are the inclusion maps. We can thus restrict each of the functors $\textswab{F},\textswab{A}:\Loc_\copyright\rightarrow\umAlg$ to $\mathcal{J}_\textbf{M}$ and obtain functors $\textswab{F}_\textbf{M},\textswab{A}_\textbf{M}:\mathcal{J}_\textbf{M}\rightarrow\umAlg.$
The universal algebras $\textswab{F}_u\textbf{M}$ and $\textswab{A}_u\textbf{M}$ are now precisely the universal objects of the \emph{colimits} (see \cite[Sec.2.5]{Par70}, \cite[Sec.2.6]{Bor94} or \cite[Sec.III.3]{McL98} for this categorical notion) for the functors $\textswab{F}_\textbf{M}$ and $\textswab{A}_\textbf{M}$ but viewed as functors $\textswab{F}_\textbf{M},\textswab{A}_\textbf{M}:\mathcal{J}_\textbf{M}\rightarrow\uAlg$. 
Here, it is crucial to drop the restriction to monic morphisms, because $\uAlg$ is \emph{cocomplete}, i.e. the colimit for any functor from any small category to $\uAlg$ always exists, while $\umAlg$ is not; in fact, the colimits for $\textswab{F}_\textbf{M}$ and $\textswab{A}_\textbf{M}$ do not exist in $\umAlg$ for general $\Mb$. This justifies the use of the term `universal'. At this point, we get the functorial property of $\textswab{F}_u,\textswab{A}_u:\Loc\rightarrow\uAlg$ for free because they necessarily turn out to be the \emph{left Kan extensions} (see \cite[Sec.3.7]{Bor94} or \cite[Chap.X]{McL98}) of $\textswab{F},\textswab{A}:\Loc_\copyright\rightarrow\uAlg$ (again, one must work in
$\uAlg$ rather than $\umAlg$). Hence, from this categorical point of view, the universal theories of the quantised free Maxwell field are highly distinguished extensions of the theories on contractible spacetimes.

The notion of a colimit and a left Kan extension also make sense for the categories $\pSymplK$, $\mpSymplK$ and $\SymplK$, but none of these three categories is cocomplete. However, it can be shown that the functors $\mathcal{F}_\textbf{M},\mathcal{A}_\textbf{M}:\mathcal{J}_\textbf{M}\rightarrow\pSymplK$ have colimits 
whose universal objects are precisely  $\mathcal{F}_u\textbf{M}$ and  $\mathcal{A}_u\textbf{M}$ respectively, and
that $\mathcal{F}_u,\mathcal{A}_u:\Loc\rightarrow\pSymplK$ are the left Kan extensions of $\mathcal{F},\mathcal{A}:\Loc_\copyright\rightarrow\pSymplK$.
Moreover, the relations $\mathcal{Q}\left(\mathcal{F}_u\textbf{M}\right)
=\textswab{F}_u\textbf{M}$ and $\mathcal{Q}\left(\mathcal{A}_u\textbf{M}\right)=
\textswab{A}_u\textbf{M}$ can be understood as special cases of a
general result. Although the colimits for $\mathcal{F}_\textbf{M},\mathcal{A}_\textbf{M}:\mathcal{J}_\textbf{M}\rightarrow\mpSymplK$ (or $\SymplK$) do not exist,  
the non-existence of colimits does not rule out the existence of left Kan extensions and it would be indeed interesting to know if $\textswab{F},\textswab{A}:\Loc_\copyright\rightarrow\umAlg$ and $\mathcal{F},\mathcal{A}:\Loc_\copyright\rightarrow\pSymplK$ (or $\SymplK$) have left Kan extensions in $\umAlg$ and $\pSymplK$ (or $\SymplK$). If they do exist, the resulting
theories would be distinguished as the minimal locally covariant extensions of the theory
on contractible spacetimes; while we have not reached a conclusion on the question of existence,
it can however be shown that {\em if} these extensions exist, they would coincide with the reduced theories.

\subsection{Maxwell theories and SPASs}\label{SPASs}
A foundational problem for physics in curved
spacetimes is to understand how a theory should be formulated such that its physical content is preserved across the various spacetimes on which it is defined; i.e., so that it represents the same physics in all spacetimes (SPASs) \cite{SPASs12}. This touches on what is actually meant by the physical content of a theory and it is not easy to make this mathematically precise. Hence, there might be more than one or even no satisfactory notion of SPASs at all. 

In \cite{SPASs12}, this problem was addressed as follows. Any putative
notion of SPASs can be represented by a class of locally covariant theories -- those conforming to the notion in question. One can then assert axioms
for what a good notion of SPASs should be as restrictions on such classes of theories. In particular, suppose one has two theories $\mathcal{A}$, $\mathcal{B}$, in a class 
$\mathfrak{T}$, each of which is supposed to represent the same physics 
in all spacetimes according to a common notion. If there is at least
one spacetime in which theories $\mathcal{A}$ and $\mathcal{B}$ coincide,  then it
seems natural to demand that they should coincide in all spacetimes. 

This idea was implemented mathematically for the case in which theory $\mathcal{A}$ is a subtheory of $\mathcal{B}$: 
A class of theories $\mathfrak{T}$ is said to have the \emph{SPASs property} if and only if whenever $\mathcal{A},\mathcal{B}:\Loc\to\Phys$ are LCTs in $\mathfrak{T}$ and $\eta:\mathcal{A}\dot{\to}\mathcal{B}$ is a partial natural isomorphism (i.e., at least one of its components is an isomorphism), then $\eta$ is a natural isomorphism. 
It was pointed out in \cite{SPASs12} that the collection of all locally covariant quantum field theories from $\Loc$ to $\Phys$ (for 
rather general choices of $\Phys$, including $\umAlg$ for example) does not have the SPASs property, while the class of dynamically local
theories does. It was also noted that one might wish to consider other implementations of the underlying idea.

The models studied in this paper provide a new viewpoint on this issue. The theories $\mathcal{F}_u$
and $\widetilde{\mathcal{F}}$ (resp., $\textswab{F}_u$ and $\widetilde{\textswab{F}}$) coincide on all spacetimes with
trivial second de Rham cohomology. To be specific, let 
$\Loc_{2}$ be the full subcategory of $\Loc$ formed by 
the spacetimes $\Mb$ with $H^2_{dR}(M)=0$, and let $K:\Loc_{2}\to \Loc$ be the inclusion functor. Then
there are natural isomorphisms
$\mathcal{F}_u\circ K\stackrel{\cdot}{\simeq}\widetilde{\mathcal{F}}\circ K$ and
$\textswab{F}_u\circ K\stackrel{\cdot}{\simeq}\widetilde{\textswab{F}}\circ K$. However, the theories
are not equivalent on $\Loc$ and it is evidently not tenable
to regard both the universal and reduced theories as 
each representing the same physics in all spacetimes according
to a common notion. 

As far as we are aware, there is no way of embedding the
reduced theories as subtheories of their universal cousins.\footnote{
In any spacetime $\Mb$ one can find a monic from
$\widetilde{\mathcal{F}}\Mb$ to $\mathcal{F}_u\Mb$, e.g.,
$\L \omega\R\mapsto \sum_\alpha [\chi_\alpha \omega]$, 
where $\chi_\alpha$ is a partition of unity subordinate to a
covering by contractible globally hyperbolic subsets; the problem
is that such maps are not generally unique and (lacking a 
global Hodge theory for $\Loc$) there is no  
natural choice.} However, it would be 
natural to regard the universal theories as {\em extensions}
of the reduced ones. In the classical theory we have
a short left exact sequence 
\begin{flalign*}
0 \stackrel{\cdot}{\longrightarrow} \rad \textswab{w}_u \xlongrightarrow[m]{\cdot} \mathcal{F}_u \xlongrightarrow[e]{\cdot}
\widetilde{\mathcal{F}},
\end{flalign*}
of functors from $\Loc$ to $\pSymplK$, where all components of $e$ are epimorphisms. Here,  $0$ denotes
the constant functor returning the zero (complexified if $\mathbb{K}=\mathbb{C}$) pre-symplectic space and
$\rad \textswab{w}_u$ is the functor assigning the radical $\rad \textswab{w}_{u\,\Mb}$ 
(equipped with the zero pre-symplectic form) to each $\Mb\in\Loc$,
and with morphisms obtained by restriction from $\mathcal{F}_u$. 
The components of $m$, which are the inclusion morphisms of
$\rad \textswab{w}_{u\,\Mb}$ as a subobject of $\mathcal{F}_u$ are necessarily monic. (As $\pSymplK$ lacks a zero object, it is not possible to 
write a short exact sequence, and we have to insist on $e$ being
epic separately.) Applying the quantisation functor, we obtain a 
similar short left exact sequence in $\uAlg$. 
In general, 
we could consider any sequence
$\mathcal{C} \xlongrightarrow[m]{\cdot} \mathcal{B} \xlongrightarrow[e]{\cdot}
\mathcal{A}$ with monic $m$ and epic $e$ as indicating that $\mathcal{B}$ is an extension of
$\mathcal{A}$ (by $\mathcal{C}$), where 
$\mathcal{A},\mathcal{B},\mathcal{C}:\Loc\to\Phys$  
(for these purposes, we would allow $\Phys$ to admit non-monic
morphisms). One may then formulate a version of the SPASs property to
cover extensions: a class $\mathfrak{T}$ of theories $\Loc\to\Phys$
has the {\em SPASs property for extensions} if, whenever $\mathcal{A},\mathcal{B}\in\mathfrak{T}$ and $\mathcal{B}$ is an extension of $\mathcal{A}$ so that $e$ is a partial natural isomorphism, then
$e$ is a natural isomorphism. It would be very interesting to know
whether the class of dynamically local theories satisfies this
version of SPASs in addition to the subtheory version studied in~\cite{SPASs12}. Our results on the Maxwell theories studied
here are certainly consistent with a positive answer to that question.


\section*{Acknowledgements}
B.L. is indebted to the Department of Mathematics, University of York, for financial support by means of a Teaching Studentship. We thank 
Ko Sanders, Alex Schenkel and Daniel Siemssen for useful comments.
{\small 
\bibliographystyle{amsalpha}
\providecommand{\bysame}{\leavevmode\hbox to3em{\hrulefill}\thinspace}
\providecommand{\MR}{\relax\ifhmode\unskip\space\fi MR }
\providecommand{\MRhref}[2]{%
  \href{http://www.ams.org/mathscinet-getitem?mr=#1}{#2}
}
\providecommand{\href}[2]{#2}
}
\end{document}